\documentclass [12pt]{article}
\usepackage {graphicx}
\usepackage{bm}
\usepackage{booktabs}
\usepackage{multirow}
\usepackage{amsmath}
\usepackage{amssymb}
\usepackage{cite}
\usepackage{amsfonts,latexsym} % to use \mathbb{} etc
\usepackage{dsfont}  % to use \mathrm{}

\newtheorem{theo}{Theorem}[section]
\newtheorem{theorem}{Theorem}[section]

\newtheorem{proof}[theo]{Proof}

\newtheorem{definition}[theo]{Definition}
\newtheorem{model ass}[theo]{Model Assumptions}
\newtheorem{model_assumptions}[theo]{Model Assumptions}

\newtheorem{example}[theo]{Example}
\newtheorem{remark}[theo]{Remarks}

\renewcommand{\vec}{\bm}

\begin{document}

\title{Loss Distribution Approach for Operational Risk Capital Modelling under Basel II: Combining Different Data Sources for Risk Estimation}

\author{\bf{Pavel V.~Shevchenko} \\
\footnotesize{(corresponding author)}\\
\small{CSIRO Mathematics, Informatics and Statistics, Australia} \\
\small{School of Mathematics and Statistics, The University of New South Wales, Australia}\\
\small{Locked Bag 17, North Ryde, NSW, 1670, Australia; e-mail: Pavel.Shevchenko@csiro.au}\\
\\
\bf{Gareth W.~Peters}\\
\small{Department of Statistical Science, University College London}\\
\small{CSIRO Mathematics, Informatics and Statistics, Australia;
email: gareth.peters@ucl.ac.uk}}

\date{\small{Draft, this version: 10 March 2013}}

\maketitle

\begin{abstract}
\noindent The management of operational risk in the banking industry
has undergone significant changes over the last decade due to
substantial changes in operational risk environment. Globalization,
deregulation, the use of complex financial products and changes in
information technology have resulted in exposure to new risks very
different from market and credit risks. In response, Basel Committee
for banking Supervision has developed a  regulatory  framework,
referred to as Basel II, that introduced operational risk category
and corresponding capital requirements. Over the past five years,
major banks in most parts of the world have received accreditation
under the Basel II Advanced Measurement Approach (AMA) by adopting
the loss distribution approach (LDA) despite there being a number of
unresolved methodological challenges in its implementation.
Different approaches and methods are still under hot debate. In this
paper, we review methods proposed in the literature for combining
different data sources (internal data, external data and scenario
analysis) which is one of the regulatory requirement for AMA.

~

\end{abstract}

\section{Operational Risk under Basel II}\label{intro_sec} The management of
operational risk in the banking industry has undergone significant
changes over the last decade due to substantial changes in
operational risk environment. Globalization, deregulation, the use
of complex financial products and changes in information technology
have resulted in exposure to new risks very different from market
and credit risks. In response, Basel Committee for banking
Supervision has developed a  regulatory  framework, referred to as
Basel II  \cite{BaselII06}, that introduced operational risk
(OpRisk) category and corresponding capital requirements against
OpRisk losses. OpRisk is defined by Basel II \cite[p.144]{BaselII06}
as: ``\emph{the risk of loss resulting from inadequate or failed
internal processes, people and systems or from external events. This
definition includes legal risk, but excludes strategic and
reputational risk.}'' Similar regulatory requirements for the
insurance industry are referred to as Solvency 2. A conceptual
difference between OpRisk and market/credit risk is that it
represents a downside risk with no upside potential.

OpRisk is significant in many financial institutions. Examples of
extremely large OpRisk losses are: Barings Bank in 1995 when the
actions of one rogue trader caused a bankruptcy as a result of GBP
1.3 billion derivative trading loss; Enron bankruptcy in 2001
considered as a result of actions of its executives with USD 2.2
billion loss; and Soci\'{e}t\'{e} G\'{e}n\'{e}rale losses of Euro
4.9 billion in 2008 due to unauthorized
 trades. In 2012, a capital against OpRisk in major Australian banks is about AUD 1.8-2.5 billion (8-10\% of the total capital). Under the
  Basel II framework, three approaches can be used to quantify the OpRisk
annual capital charge $C$, see \cite[pp.144-148]{BaselII06}.
\begin{itemize}
\item{\textbf{The Basic Indicator Approach}: $C = \alpha \frac{1}{n} \sum\nolimits_{j = 1}^3 \max({GI(j)},0)$,
where $GI(j),j=1,\ldots,3$ are the annual gross incomes over the
previous three years, $n$ is the number of years with positive gross
income, and $\alpha = 0.15$.}

\item{\textbf{The Standardised Approach}: $C = \frac{1}{3}\sum\nolimits_{j = 1}^3 {\max [\sum\nolimits_{i = 1}^8 {\beta _i
GI_i (j)} ,0]} $, where $\beta _i $, $i = 1,\ldots,8$ are the
factors for eight business lines (BL) listed in Table
\ref{BURT_table} and $GI_i (j)$, $j = 1,2,3$ are the annual gross
incomes of the $i$-th BL in the previous three years.}

\item{\textbf{The Advanced Measurement Approaches} (AMA): a bank can calculate
the capital charge using internally developed model subject to
regulatory approval.}

\end{itemize}

\noindent A bank intending to use the AMA should demonstrate
accuracy of the internal models within the Basel II risk cells
(eight business lines times seven risk types, see Table
\ref{BURT_table}) relevant to the bank and satisfy some criteria,
see \cite[pp.148-156]{BaselII06}, including:

\begin{itemize}
\item{The use of the internal data, relevant external data, scenario
analysis and factors reflecting the business environment and
internal control systems;}

\item{The risk measure used for capital charge should correspond to
the 99.9{\%} confidence level for a one-year holding period;}

\item{Diversification benefits are allowed if dependence modeling is
approved by a regulator;}

\item{Capital reduction due to insurance is capped by 20{\%}.}
\end{itemize}

The intention of AMA is to provide incentive to a bank to invest
into development of a sound OpRisk practices and risk management.
The capital reserves under AMA (when compared to other approaches)
will be more relevant to the actual risk profile of a bank. It is
expected that the capital from the AMA is lower than the capital
calculated under the Standardised Approach (some regulators are
setting a limit on this reduction, e.g. 25\%). The regulatory
accreditation for AMA indicates to a market that a bank has
developed a sound risk management practice.

\begin{table}[!htbp]
{\footnotesize{
\begin{tabular}
{p{0.45\textwidth}p{0.505\textwidth}} \toprule \textbf{\quad Basel
II business lines (BL)}&
\textbf{\quad Basel II event types (ET)} \\
\cmidrule(r){1-1}\cmidrule(l){2-2}
\begin{itemize}
\item Corporate finance ($\beta_{1}=0.18$)
\item Trading {\&} Sales ($\beta_{2}=0.18$)
\item Retail banking ($\beta_{3}=0.12$)
\item Commercial banking ($\beta_{4}=0.15$)
\item Payment {\&} Settlement ($\beta_{5}=0.18$)
\item Agency Services ($\beta_{6}=0.15$)
\item Asset management ($\beta_{7}=0.12$)
\item Retail brokerage ($\beta_{8}=0.12$)
\end{itemize}
&
\begin{itemize}
\item Internal fraud
\item External fraud
\item{Employment practices and \par workplace safety}
\item Clients, products and business practices
\item Damage to physical assets
\item Business disruption and system failures
\item{Execution, delivery and \par process management}
\end{itemize}\\
\bottomrule
\end{tabular}
}} \caption{\small{Basel II business lines and event types. $\beta
_1 ,\ldots,\beta _8$ are the business line factors used in the Basel
II Standardised Approach.}} \label{BURT_table}
\end{table}

\begin{remark} While the regulatory capital for operational risk is
based on the 99.9\% confidence level over a one year period,
economic capital used by banks is often higher; some banks use the
99.95\%-99.98\% confidence levels.
\end{remark}

\noindent A popular method under the AMA is the loss distribution
approach (LDA). Under the LDA, banks quantify distributions for
frequency and severity of OpRisk losses for each risk cell (business
line/event type) over a one-year time horizon. The banks can use
their own risk cell structure but must be able to map the losses to
the Basel II risk cells. There are various quantitative aspects of
the LDA modeling discussed in several books
\cite{King01,Cruz02,Cruz04,Panjer06, McFrEm05, ChRaFa07} and various
papers, e.g. \cite{ChEmNe06,FrMoRo04,AuKl06} to mention a few. The
commonly used LDA model for calculating the total annual loss $Z
(t)$ in a bank (occurring in the years $t = 1,2,\ldots)$ can be
formulated as
\begin{equation}
\label{LDAmodel_eq} Z (t) = \sum\limits_{j = 1}^J {Z_j (t)} ;\quad
Z_j (t) = \sum\limits_{i = 1}^{N_j (t)} {X_i^{(j)} (t)} .
\end{equation}
\noindent Here, the annual loss $Z_j (t)$ in risk cell $j$ is
modeled as a compound process over one year with the frequency
(annual number of events) $N_j (t)$ implied by a counting process
(e.g. Poisson process) and random severities $X_i^{(j)} (t)$, $i =
1,\ldots,N_j (t)$. Estimation of the annual loss distribution by
modeling frequency and severity of losses is a well-known actuarial
technique used to model solvency requirements for the insurance
industry, see e.g. \cite{KlPaWi98,Sandstrom06,WuMe08}. Then the
capital is defined as the 0.999 Value at Risk (VaR) which is the
quantile of the distribution for the next year total annual loss
$Z(T + 1)$:
\begin{equation}
\label{VaRdef_eq} VaR_q [Z (T + 1)] = F_{Z (T + 1)}^{ - 1} (q) =
\inf \{z:\Pr [Z (T + 1) > z] \le 1 - q\}
\end{equation}
\noindent at the level $q = 0.999$. Here, index $T$+1 refers to the
next year and notation $F_Y^{ - 1} (q)$ denotes the inverse
distribution of a random variable  $Y$. The capital can be
calculated as the difference between the 0.999 VaR and expected loss
if the bank can demonstrate that the expected loss is adequately
captured through other provisions. If correlation assumptions can
not be validated between some groups of risks (e.g. between business
lines) then the capital should be calculated as the sum of the 0.999
VaRs over these groups. This is equivalent to the assumption of
perfect positive dependence between annual losses of these groups.
However, it is important to note that the sum of VaRs across risks
is not most conservative estimate of the total VaR. In principle,
the upper conservative bound  can be larger; see Embrechts et al
\cite{EmNeWu09} and Embrechts et al \cite{EmLaWu09}. This is often
the case for heavy tailed distributions (with the tail decay slower
than the exponential) and large quantiles.

The major problem in OpRisk is a lack of quality data that makes it
difficult for advanced research in the area. In past, most banks did
not collect OpRisk data -- it was not required while the cost of
collection is significant. Moreover, indirect OpRisk losses cannot
be measured accurately. Also the duration of OpRisk events can be
substantial and evaluation of the impact of the event can take
years.

Over the past five years, major banks in most parts of the world
have received accreditation under the Basel II AMA by adopting the
LDA despite there being a number of unresolved methodological
challenges in its implementation. Different approaches and methods
are still under hot debate. One of the unresolved challenges is
combining internal data with external data and scenario analysis
required by Basel II. In this paper, we review some methods proposed
in the literature to combine different data sources for OpRisk
capital modelling. Other challenges not discussed in this paper
include modelling dependence between risks, handling data
truncation, modelling heavy tailed severities, and estimation of the
frequency and severity distributions; for these issues, the readers
are refereed to Panjer \cite{Panjer06} or Shevchenko
\cite{shevchenko2011modelling}.

The paper is organised as follows. Section \ref{data_sec} describes
the requirements for the data that should be collected and used for
Basel II AMA.  Combining different data sources using ad-hoc and
Baysian methods are considered in Sections
\ref{adhoc_sec}--\ref{Chapter_combining_3sources_section}. Other
methods of combining, non-parametric Bayesian method via Dirichlet
process and Dempster's combining rule are considered in Section
\ref{nonparametricBayesian_sec} and Section \ref{DempsetShafer_sec}
respectively. To avoid confusion in description of mathematical
concepts we follow a standard statistical notation denoting random
variables by upper case symbols and their realisations by lower case
symbols.

\section{Data Sources}
\label{data_sec} Basel II specifies requirement for the data that
should be collected and used for AMA. In brief, a bank should have
internal data, external data and expert opinion data. In addition,
internal control indicators and factors affecting the businesses
should be used. A bank's methodology must capture key business
environment and internal control factors affecting OpRisk. These
factors should help to make forward-looking estimation, account for
the quality of the controls and operating environments, and align
capital assessments with risk management objectives.

The intention of the use of several data sources is to develop a
model based on the largest possible dataset to increase the accuracy
and stability of the capital estimate. Development and maintenance
of OpRisk databases is a difficult and challenging task. Some of the
main features of the required data are summarized as follows.

\subsection{Internal data} The internal data should be collected
over a minimum five year period to be used for capital charge
calculations (when the bank starts the AMA, a three-year period is
acceptable). Due to a short observation period, typically, the
internal data for many risk cells contain few (or none) high impact
low frequency losses. A bank must be able to map its historical
internal loss data into the relevant Basel II risk cells in Table
\ref{BURT_table}. The data must capture all material activities and
exposures from all appropriate sub-systems and geographic locations.
A bank can have an appropriate reporting threshold for internal data
collection, typically of the order of Euro 10,000. Aside from
information on gross loss amounts, a bank should collect information
about the date of the event, any recoveries of gross loss amounts,
as well as some descriptive information about the drivers of the
loss event.

\subsection{External data} A bank's OpRisk measurement system must
use relevant external data. These data should include data on actual
loss amounts, information on the scale of business operations where
the event occurred, and information on the causes and circumstances
of the loss events. Industry data are available through external
databases from vendors (e.g. Algo OpData provides publicly reported
OpRisk losses above USD 1 million) and consortia of banks (e.g. ORX
provides OpRisk losses above Euro 20,000 reported by ORX members).
The external data are difficult to use directly due to different
volumes and other factors. Moreover, the data have a survival bias
as typically the data of all collapsed companies are not available.
Several Loss Data Collection Exercises (LDCE) for historical OpRisk
losses over many institutions were conducted and their analyses
reported in the literature. In this respect, two papers are of high
importance: \cite{Moscadelli04} analysing 2002 LDCE and
\cite{DuPe06} analysing 2004 LDCE where the data were mainly above
Euro 10,000 and USD 10,000 respectively. To show the severity and
frequency of operational losses, Table
\ref{Chapter_intro_BURT2004LDCEtable} presents a data summary for
2004 LDCE conducted by US Federal bank and Thrift Regulatory
agencies in 2004 for US banks. Here, twenty three US banks provided
data for about 1.5 million losses totaling USD 25.9 billion. It is
easy to see that frequencies and severities of losses are very
different across risk cells, though some of the cells have very few
and small losses.

\subsection{Scenario Analysis} A bank must use scenario analysis in
conjunction with external data to evaluate its exposure to
high-severity events. Scenario analysis is a process undertaken by
experienced business managers and risk management experts to
identify risks, analyse past internal/external events, consider
current and planned controls in the banks; etc. It may involve:
workshops to identify weaknesses, strengths and other factors;
opinions on the impact and likelihood of losses; opinions on sample
characteristics or distribution parameters of the potential losses.
As a result some rough quantitative assessment of risk frequency and
severity distributions can be obtained. Scenario analysis is very
subjective and should be combined with the actual loss data. In
addition, it should be used for stress testing, e.g. to assess the
impact of potential losses arising from multiple simultaneous loss
events.

Expert opinions on potential losses and corresponding probabilities
are often expressed using opinion on the distribution parameter;
opinions on the number of losses with the amount to be within some
ranges; separate opinions on the frequency of the losses and
quantiles of the severity; opinion on how often the loss exceeding
some level may occur. Expert elicitation is certainly one of the
challenges in OpRisk because many managers and employees may not
have a sound knowledge of statistics and probability theory. This
may lead to misleading and misunderstanding. It is important that
questions answered by experts are simple and well understood by
respondents. There are psychological aspects involved. There is a
vast literature on expert elicitation published by statisticians,
especially in areas such as security and ecology. For a good review,
see O'Hagan \cite{OHagan06}. However, published studies on the use
of expert elicitation for OpRisk LDA are scarce. Among the few are
Frachot et al \cite{FrMoRo04}; Alderweireld et al \cite{AlGaLe06};
Steinhoff and Baule \cite{StBa06}; and Peters and H\"ubner
\cite{PeHu09}. These studies suggest that questions on \emph{``how
often the loss exceeding some level may occur"} are well understood
by OpRisk experts. Here, experts express the opinion that a loss of
amount $L$ or higher is expected to occur every $d$ years. A
recently proposed framework that incorporates scenario analysis into
OpRisk modeling was proposed in Ergashev \cite{Ergashev12}, where
the basis for the framework is the idea that only worst-case
scenarios contain valuable information about the tail behavior of
operational losses.

\begin{remark} One of the problems with the combining external data and scenario analysis is that external data are collected for Basel II risk cells
while scenario analysis is done at the loss process level.
\end{remark}

\subsection{A Note on Data Sufficiency.} Empirical estimation
of the annual loss 0.999 quantile, using observed losses only, is
impossible in practice. It is instructive to calculate the number of
data points needed to estimate the 0.999 quantile empirically within
the desired accuracy. Assume that independent data points
$X_1,\ldots,X_n$ with common density $f(x)$ have been observed. Then
the quantile $q_\alpha$ at confidence level $\alpha$ is estimated
empirically as $\widehat{Q}_\alpha=\widetilde{X}_{\left\lfloor
{n\alpha} \right\rfloor +1}$, where $\widetilde {\vec X}$ is the
data sample $\vec{X}$ sorted into the ascending order. The standard
deviation of this empirical estimate is
\begin{equation}
\mathrm{stdev}[\widehat{Q}_\alpha]=\frac{\sqrt{\alpha(1-\alpha)}}{f(q_\alpha)\sqrt{n}};
\end{equation}
\noindent see Glasserman \cite[section 9.1.2, p. 490]{Glasserman04}.
Thus, to calculate the quantile within relative error
$\varepsilon=2\times\mathrm{stdev}[\widehat{Q}_\alpha]/q_\alpha$, we
need
\begin{equation}
\label{Chapter_EVT_ndataTogetAccuracy_eq}
n=\frac{4\alpha(1-\alpha)}{\varepsilon^2 (f(q_\alpha)q_\alpha)^2}
\end{equation}
\noindent observations. Suppose that the data are from the lognormal
distribution $\mathcal{LN}(\mu=0,\sigma=2)$. Then using formula
(\ref{Chapter_EVT_ndataTogetAccuracy_eq}), we obtain that
$n=140,986$ observations are required to achieve $10\%$ accuracy
($\varepsilon=0.1$) in the 0.999 quantile estimate. In the case of
$n=1,000$ data points, we get $\varepsilon=1.18$, that is, the
uncertainty is larger than the quantile we estimate. Moreover,
according to the regulatory requirements, the 0.999 quantile of the
annual loss (rather than 0.999 quantile of the severity) should be
estimated. OpRisk losses are typically modelled by the heavy-tailed
distributions. In this case, the quantile at level $q$ of the
aggregate distributions can be approximated by the quantile of the
severity distribution at level
$$p=1-\frac{1-q}{\mathrm{E}[N]};$$

\noindent see Embrechts et al \cite[theorem 1.3.9]{EmKlMi97}. Here,
$\mathrm{E}[N]$ is the expected annual number of events. For
example, if $\mathrm{E}[N]=10$, then we obtain that the error of the
annual loss 0.999 quantile is the same as the error of the severity
quantile at the confidence level $p=0.9999$. Again, using
(\ref{Chapter_EVT_ndataTogetAccuracy_eq}) we conclude that this
would require $n\approx 10^6$ observed losses to achieve $10\%$
accuracy. If we collect annual losses then $n/\mathrm{E}[N]\approx
10^5$ annual losses should be collected to achieve the same accuracy
of $10\%$. These amounts of data are not available even from the
largest external databases and extrapolation well beyond the data is
needed. Thus parametric models must be used. For an excellent
discussion on data sufficiency in OpRisk, see Cope et al
\cite{CoAnMiUg09}.

\begin{table}[!htbp]
\caption{\small{Number of loss events (\%, top value in a cell) and
total Gross Loss (\%, bottom value in a cell) annualised per
Business Line and Event Type reported by US banks in 2004 LDCE
\cite[tables 3 and 4]{LDCE2004}. 100\% corresponds to 18,371.1
events and USD 8,643.2 million. Losses $\geq$ USD 10,000 occurring
during the period 1999-2004 in years when data capture was stable.}}
{\footnotesize{
\begin{tabular} {@{}
p{0.07\textwidth}@{\hspace{0.01\textwidth}}
p{0.08\textwidth}@{\hspace{0.01\textwidth}}
p{0.08\textwidth}@{\hspace{0.01\textwidth}}
p{0.08\textwidth}@{\hspace{0.01\textwidth}}
p{0.08\textwidth}@{\hspace{0.02\textwidth}}
p{0.08\textwidth}@{\hspace{0.01\textwidth}}
p{0.08\textwidth}@{\hspace{0.01\textwidth}}
p{0.08\textwidth}@{\hspace{0.01\textwidth}}
p{0.08\textwidth}@{\hspace{0.01\textwidth}}
p{0.08\textwidth}@{\hspace{0.03\textwidth}} p{0.08\textwidth}@{}}
%{|p{1.0cm}|p{0.9cm}|p{1.0cm}|p{0.9cm}|p{0.9cm}|p{0.9cm}|p{0.9cm}|p{1.0cm}|p{0.9cm}|p{0.9cm}||p{1.1cm}|}
\toprule & \textbf{ET(1)}& \textbf{ET(2)}& \textbf{ET(3)}&
\textbf{ET(4)}& \textbf{ET(5)}& \textbf{ET(6)}& \textbf{ET(7)}&
\textbf{Other}& \textbf{Fraud}&
\textbf{Total} \\
\midrule \multirow{2}{0.07\textwidth}{\textbf{BL(1)}}& 0.01{\%} \par
0.14{\%}& 0.01{\%} \par 0.00{\%}& 0.06{\%} \par 0.03{\%}& 0.08{\%}
\par 0.30{\%}& 0.00{\%}
\par 0.00{\%}& & 0.12{\%} \par 0.05{\%}& 0.03{\%} \par 0.01{\%}&
0.01{\%} \par 0.00{\%}&
0.3{\%} \par 0.5{\%} \\
\midrule
 \multirow{2}{0.07\textwidth}{\textbf{BL(2)}}& 0.02{\%} \par
0.10{\%}& 0.01{\%} \par \textbf{1.17{\%}}& 0.17{\%} \par 0.05{\%}&
0.19{\%} \par \textbf{4.29{\%}}& 0.03{\%} \par 0.00{\%}& 0.24{\%}
\par 0.06{\%}& \textbf{6.55{\%}} \par \textbf{2.76{\%}}& & 0.05{\%}
\par 0.15{\%}&
\textbf{7.3{\%}} \par \textbf{8.6{\%}} \\
\midrule
 \multirow{2}{0.07\textwidth}{\textbf{BL(3)}}&
\textbf{2.29{\%}}
\par 0.42{\%}& \textbf{33.85{\%}} \par \textbf{2.75{\%}}&
\textbf{3.76{\%}} \par 0.87{\%}& \textbf{4.41{\%}} \par
\textbf{4.01{\%}}& 0.56{\%} \par 0.1{\%}& 0.21{\%} \par 0.21{\%}&
\textbf{12.28{\%}} \par \textbf{3.66{\%}}& 0.69{\%} \par 0.06{\%}&
\textbf{2.10{\%}} \par 0.26{\%}&
\textbf{60.1{\%}} \par \textbf{12.3{\%}} \\
\midrule
 \multirow{2}{0.07\textwidth}{\textbf{BL(4)}}& 0.05{\%} \par
0.01{\%}& \textbf{2.64{\%}}
\par 0.70{\%}& 0.17{\%} \par 0.03{\%}& 0.36{\%} \par 0.78{\%}&
0.01{\%} \par 0.00{\%}& 0.03{\%} \par 0.00{\%}& \textbf{1.38{\%}}
\par 0.28{\%}& 0.02{\%} \par 0.00{\%}& 0.44{\%} \par 0.04{\%}&
\textbf{5.1{\%}} \par \textbf{1.8{\%}} \\
\midrule
 \multirow{2}{0.07\textwidth}{\textbf{BL(5)}}& 0.52{\%} \par
0.08{\%}& 0.44{\%} \par 0.13{\%}& 0.18{\%} \par 0.02{\%}& 0.04{\%}
\par 0.01{\%}& 0.01{\%}
\par 0.00{\%}& 0.05{\%} \par 0.02{\%}& \textbf{2.99{\%}} \par
0.28{\%}& 0.01{\%} \par 0.00{\%}& 0.23{\%} \par 0.05{\%}&
\textbf{4.5{\%}} \par 0.6{\%} \\
\midrule
 \multirow{2}{0.07\textwidth}{\textbf{BL(6)}}& 0.01{\%} \par
0.02{\%}& 0.03{\%} \par 0.01{\%}& 0.04{\%} \par 0.02{\%}& 0.31{\%}
\par 0.06{\%}& 0.01{\%}
\par 0.01{\%}& 0.14{\%} \par 0.02{\%}& \textbf{4.52{\%}} \par
0.99{\%}& & &
\textbf{5.1{\%}} \par \textbf{1.1{\%}} \\
\midrule
 \multirow{2}{0.07\textwidth}{\textbf{BL(7)}}& 0.00{\%} \par
0.00{\%}& 0.26{\%} \par 0.02{\%}& 0.10{\%} \par 0.02{\%}& 0.13{\%}
\par \textbf{2.10{\%}}& 0.00{\%} \par 0.00{\%}& 0.04{\%} \par
0.01{\%}& \textbf{1.82{\%}}
\par 0.38{\%}& & 0.09{\%} \par 0.01{\%}&
\textbf{2.4{\%}} \par \textbf{2.5{\%}} \\
\midrule
 \multirow{2}{*}{\textbf{BL(8)}}& 0.06{\%} \par 0.03{\%}&
0.10{\%}
\par 0.02{\%}& \textbf{1.38{\%}} \par 0.33{\%}& \textbf{3.30{\%}}
\par 0.94{\%}& & 0.01{\%} \par 0.00{\%}& \textbf{2.20{\%}} \par
0.25{\%}& & 0.20{\%} \par 0.07{\%}&
\textbf{7.3{\%}} \par \textbf{1.6{\%}} \\
\midrule \multirow{2}{0.07\textwidth}{\textbf{Other}}& 0.42{\%} \par
0.1{\%}& \textbf{1.66{\%}} \par 0.3{\%}& \textbf{1.75{\%}} \par
0.34{\%}& 0.40{\%} \par \textbf{67.34{\%}}& 0.12{\%} \par
\textbf{1.28{\%}}& 0.02{\%} \par 0.44{\%}& \textbf{3.45{\%}} \par
0.98{\%}& 0.07{\%}
\par 0.05{\%}& 0.08{\%} \par 0.01{\%}&
\textbf{8.0{\%}} \par \textbf{70.8{\%}} \\
\midrule\midrule
 \multirow{2}{*}{\textbf{Total}} \par  &
\textbf{3.40{\%}}
\par 0.9{\%}& \textbf{39.0{\%}} \par \textbf{5.1{\%}}&
\textbf{7.6{\%}} \par \textbf{1.7{\%}}& \textbf{9.2{\%}} \par
\textbf{79.8{\%}}& 0.7{\%}
\par \textbf{1.4{\%}}& 0.7{\%} \par 0.8{\%}& \textbf{35.3{\%}} \par
\textbf{9.6{\%}}& 0.8{\%} \par 0.1{\%}& \textbf{3.2{\%}} \par
0.6{\%}&
100.0{\%} \par 100.0{\%} \\
\bottomrule
\end{tabular}
}}
\label{Chapter_intro_BURT2004LDCEtable}
\end{table}

\subsection{Combining different data sources} Estimation of
low-frequency/high-severity risks cannot be done using historically
observed losses from one bank only. It is just not enough data to
estimate high quantiles of the risk distribution. Other sources of
information that can be used to improve risk estimates and are
required by the Basel II for OpRisk AMA are internal data, relevant
external data, scenario analysis and factors reflecting the business
environment and internal control systems. Specifically, Basel II AMA
includes the following requirement\footnote{The original text is
available free of charge on the BIS website
www.BIS.org/bcbs/publ.htm.} \cite[p. 152]{BaselII06}: \emph{``Any
operational risk measurement system must have certain key features
to meet the supervisory soundness standard set out in this section.
These elements must include the use of internal data, relevant
external data, scenario analysis\index{scenario analysis} and
factors reflecting the business environment and internal control
systems."}

Combining these different data sources for model estimation is
certainly one of the main challenges in OpRisk. Conceptually, the
following ways have been proposed to process different data sources
of information:
\begin{itemize}
\item numerous ad-hoc procedures;
\item parametric and nonparametric Bayesian methods; and
\item general non-probabilistic methods such as Dempster-Shafer
theory.
\end{itemize}

These methods are presented in the following sections. Methods of
credibility theory, closely related to Bayesian method are not
considered in this paper; for applications in the context of OpRisk,
see \cite{BuShWu07}. For application of Bayesian networks for
OpRisk, the reader is referred to \cite{NeFeTa05} and
\cite{NeHaAn09}. Another challenge in OpRisk related to scaling of
external data with respect to bank factors such as total assets,
number of employees, etc is not reviewed in this paper; interested
reader is referred to a recent study Ganegoda and Evans
\cite{GanegodaEvans2013}.

\section{Ad-hoc Combining}\label{adhoc_sec}
Often in practice, accounting for factors reflecting the business
environment and internal control systems is achieved via scaling of
data. Then ad-hoc procedures are used to combine internal data,
external data and expert opinions. For example:

\begin{itemize}
\item Fit the severity distribution to the combined samples of
internal and external data and fit the frequency distribution using
internal data only.

\item Estimate the Poisson annual intensity for the frequency
distribution as $w\lambda _{int} + (1 - w)\lambda _{ext} $, where
the intensities $\lambda _{ext} $ and $\lambda _{int} $ are implied
by the external and internal data respectively, using expert
specified weight $w$\index{weight!combining data}.

\item Estimate the severity distribution as a mixture
$$w_1 F_{SA} (x)
+ w_2 F_I (x) + (1 - w_1 - w_2 )F_E (x),$$ \noindent where $F_{SA}
(x)$, $F_I (x)$ and $F_E (x)$ are the distributions identified by
scenario analysis, internal data and external data respectively,
using expert specified weights $w_1 $ and $w_2 $.

\item Apply the \emph{minimum variance principle}, where the
combined estimator is a linear combination of the individual
estimators obtained from internal data, external data and expert
opinion separately with the weights chosen to minimize the variance
of the combined estimator.
\end{itemize}

Probably the easiest to use and most flexible procedure is the
minimum variance principle. The rationale behind the principle is as
follows. Consider two unbiased independent estimators $\widehat
{\Theta }^{(1)}$ and $\widehat{\Theta }^{(2)}$ for parameter
$\theta$, i.e. $\mathrm{E}[\widehat {\Theta }^{(k)}\mbox{]} = \theta
$ and $\mathrm{Var}[\widehat {\Theta }^{(k)}] = \sigma_k^2 $, $k =
1,2$. Then the combined unbiased linear estimator and its variance
are
\begin{equation}
\widehat {\Theta }_{tot} = w_1 \widehat {\Theta }^{(1)} + w_2
\widehat {\Theta }^{(2)},\quad w_1 + w_2 = 1,
\end{equation}
\begin{equation}
 \mathrm{Var}[\widehat {\Theta }_{tot} ] = w_1^2 \sigma _1^2 + (1 - w_1 )^2\sigma _2^2 .
\end{equation}
\noindent It is easy to find the weights\index{weight!minimum
variance} minimising $\mathrm{Var}[\widehat {\Theta }_{tot}]$:
${w}_1 = {\sigma _2^2 }/({\sigma _1^2 + \sigma _2^2 })$ and ${w}_2 =
{\sigma _1^2 }/({\sigma _1^2 + \sigma _2^2 })$. The weights behave
as expected in practice. In particular, ${w}_1 \to 1$ if $\sigma
_1^2 / \sigma _2^2 \to 0$ ($\sigma _1^2 / \sigma _2^2 $ is the
uncertainty of the estimator $\widehat{\Theta }^{(1)}$ over the
uncertainty of $\widehat{\Theta }^{(2)})$ and ${w}_1 \to 0$ if
$\sigma _2^2 / \sigma _1^2 \to 0$. This method can easily be
extended to combine three or more estimators using the following
theorem.

\begin{theorem}[Minimum variance estimator]
\label{Chapter_combining_MinVarEstimator_Theorem} Assume that we
have $\widehat{\Theta }^{(i)}$, $i=1,2,\ldots,K$ unbiased and
independent estimators of $\theta$ with variances
$\sigma^2_i=\mathrm{Var}[\Theta^{(i)}]$. Then the linear estimator
\begin{equation*}
\widehat {\Theta }_{tot} = w_1 \widehat {\Theta }^{(1)} + \cdots +
w_K \widehat {\Theta }^{(K)},
\end{equation*}
is unbiased and has a minimum variance if $
w_i=({1/\sigma^2_i})/{\sum_{k=1}^{K}(1/\sigma_k^2)}$. In this case,
$w_1 + \cdots + w_K = 1$ and
$$\mathrm{Var}[\widehat{\Theta}_{tot}]=\left(\sum_{k=1}^K\frac{1}{\sigma_k^2}
\right)^{-1}.$$
\end{theorem}

This result is well known, for a proof, see e.g. Shevchenko
\cite[exercise problem 4.1]{shevchenko2011modelling}. It is a simple
exercise to extend the above principle to the case of unbiased
estimators with known linear correlations. Heuristically, minimum
variance principle can be applied to almost any quantity, including
a distribution parameter or distribution characteristic such as
mean, variance or quantile. The assumption that the estimators are
unbiased estimators for $\theta$ is probably reasonable when
combining estimators from different experts (or from expert and
internal data). However, it is certainly questionable if applied to
combine estimators from the external and internal data.

\section{Bayesian Method to Combine Two Data Sources}\label{Chapter_combining_twosources_sec} The Bayesian inference
method can be used to combine different data sources in a consistent
statistical framework. Consider a random vector of data ${\vec{X}} =
(X_1 ,X_2 ,\ldots,X_n )^\prime$ whose joint density, for a given
vector of parameters ${\vec \Theta}  = (\Theta _1 ,\Theta _2
,\ldots,\Theta _K )^\prime$, is $h({\vec{x}}\vert {{\vec\theta }})$.
In the Bayesian approach, both observations and parameters are
considered to be random. Then the joint density is
\begin{equation}
\label{Chapter_combining_JointDensity} h({\vec{x}},{{\vec\theta }})
= h({\vec{x}}\vert {{\vec\theta }})\pi({{\vec\theta}}) =
\pi({{\vec\theta }}\vert {\vec{x}})h({\vec{x}}),
\end{equation}
\noindent where
\begin{itemize}
\item $\pi ({{\vec\theta }})$ is the probability density
of the parameters, a so-called prior density function. Typically,
$\pi ({ {\vec\theta }})$ depends on a set of further parameters that
are called hyper-parameters\index{hyper-parameters}, omitted here
for simplicity of notation;
\item $\pi({{\vec\theta }}\vert {\vec{x}})$ is the density
of parameters given data ${\vec{X}}$, a so-called posterior density;

\item $h({\vec{x}},{{\vec \theta }})$ is the joint
density of observed data and parameters;

\item $h({\vec{x}}\vert{
{\vec\theta }})$ is the density of observations for given
parameters. This is the same as a likelihood function if considered
as a function of $\vec\theta$, i.e.
$\ell_{\vec{x}}(\vec\theta)=h({\vec{x}}\vert{ {\vec\theta }})$;

\item $h({\vec{x}})$ is a marginal density of ${\vec
{X}}$ that can be written as $ h({\vec{x}}) = \int {h({\vec{x}}\vert
{{\vec\theta }})\pi ({{\vec\theta }})d{{\vec\theta }}}$. For
simplicity of notation, we consider continuous $\pi ({{\vec \theta
}})$ only. If $\pi ({{\vec \theta }})$ is a discrete probability
function, then the integration in the above expression should be
replaced by a corresponding summation.
\end{itemize}

\subsection{Predictive distribution}\index{predictive distribution}
The objective (in the context of OpRisk) is to estimate the
predictive distribution (frequency and severity) of a future
observation $X_{n + 1} $ conditional on all available information
${\vec{X}} = (X_1 ,X_2 ,\ldots,X_n )$. Assume that conditionally,
given $\vec{\Theta}$, $X_{n + 1} $ and ${\vec{X}}$ are independent,
and $X_{n + 1} $ has a density $f(x_{n + 1} \vert {{\vec\theta }})$.
It is even common to assume that $X_1 ,X_2 ,\ldots,X_n ,X_{n + 1} $
are all conditionally independent (given $\vec\Theta$) and
identically distributed. Then the conditional density of $X_{n + 1}
$, given data $\vec{X}=\vec{x}$, is

\begin{equation}
\label{Chapter_combining_fullpredictiveDensity}
 f(x_{n + 1} \vert {\vec{x}}) = \int {f(x_{n + 1} \vert {
{\vec\theta }})\pi({{\vec\theta }}\vert {\vec{x}})d{{\vec\theta }}}
.
\end{equation}
\noindent If $X_{n + 1} $ and ${\vec{X}}$ are not independent, then
the predictive distribution should be written as
\begin{equation}
 f(x_{n + 1} \vert {\vec{x}}) = \int {f(x_{n + 1} \vert {
{\vec\theta }},\vec{x})\pi({{\vec\theta }}\vert
{\vec{x}})d{{\vec\theta }}} .
\end{equation}

\subsection{Posterior distribution.} Bayes's theorem says
that the posterior density can be calculated from
(\ref{Chapter_combining_JointDensity}) as

\begin{equation}
\label{Chapter_combining_posterior}
 \pi({{\vec\theta }}\vert {\vec{x}}) = h({\vec{x}}\vert {{\vec\theta }})\pi ({{\vec\theta }}) /
 h({\vec{x}}).
\end{equation}

\noindent Here, $h({\vec{x}})$ plays the role of a normalisation
constant. Thus the posterior distribution can be viewed as a product
of a prior knowledge with a likelihood function for observed data.
In the context of OpRisk, one can follow the following three logical
steps.

\begin{itemize}
\item The prior distribution $\pi ({{\vec\theta }})$ should be
estimated by scenario analysis (expert opinions with reference to
external data).

\item Then the prior distribution should be weighted with the observed
data using formula (\ref{Chapter_combining_posterior}) to get the
posterior distribution $\pi({{\vec \theta }}\vert {\vec{x}})$.

\item Formula (\ref{Chapter_combining_fullpredictiveDensity}) is then used to calculate the predictive
distribution of $X_{n + 1} $ given the data ${\vec{X}}$.

\end{itemize}

\begin{remark}
~
\begin{itemize}
\item Of course, the posterior density can be used to find parameter
point estimators. Typically, these are the mean, mode or median of
the posterior. The use of the posterior mean as the point parameter
estimator is optimal in a sense that the mean square error of
prediction is minimised. For more on this topic, see B\"{u}hlmann
and Gisler \cite[section 2.3]{BuGi05}. However, in the case of
OpRisk, it is more appealing to use the whole posterior to calculate
the predictive distribution
(\ref{Chapter_combining_fullpredictiveDensity}).
\item So-called conjugate
distributions, where prior and posterior distributions are of the
same type, are very useful in practice when Bayesian inference is
applied. Below we present conjugate pairs (Poisson-gamma,
lognormal-normal) that are good illustrative examples for modelling
frequencies and severities in OpRisk. Several other pairs
 can be found, for
example, in B\"{u}\-hl\-mann and Gisler \cite{BuGi05}. In all these
cases the posterior distribution parameters are easily calculated
using the prior distribution parameters and observations. In
general, the posterior should be estimated numerically using e.g.
Markov chain Monte Carlo methods, see Shevchenko \cite[chapter
2]{shevchenko2011modelling}.
\end{itemize}
\end{remark}

\subsection{Iterative Calculation}
If the data $X_1 ,X_2 ,\ldots,X_n $ are conditionally (given
$\vec\Theta={{\vec\theta }}$) independent and $X_k$ is distributed
with a density $f_k(\cdot\vert {{\vec\theta }})$, then the joint
density of the data for given ${{\vec\theta }}$ can be written as
$h({\vec{x}}\vert {{\vec\theta }}) = \prod\limits_{i = 1}^n {f_i(x_i
\vert {{\vec\theta }})} $. Denote the posterior density calculated
after $k$ observations as $\pi_k ({{\vec\theta }}\vert x_1
,\ldots,x_k )$, then using (\ref{Chapter_combining_posterior}),
observe that

\begin{eqnarray}
\label{Chapter_combining_iterativePosterior} \pi_k ({{\vec\theta
}}\vert x_1 ,\ldots,x_k ) &\propto &\pi ({{\vec\theta
}})\prod\limits_{i = 1}^k {f_i(x_i \vert {{\vec\theta }})}
\nonumber\\
&\propto& \pi_{k - 1} ({{\vec\theta }}\vert x_1 ,\ldots,x_{k - 1} )
f_k(x_k \vert {{\vec\theta }}).
\end{eqnarray}

It is easy to see from (\ref{Chapter_combining_iterativePosterior}),
that the updating procedure which calculates the posteriors from
priors can be done iteratively. Only the posterior distribution
calculated after $k$-1 observations and the $k$-th observation are
needed to calculate the posterior distribution after $k$
observations. Thus the loss history over many years is not required,
making the model easier to understand and manage, and allowing
experts to adjust the priors at every step. Formally, the posterior
distribution calculated after $k$-1 observations can be treated as a
prior distribution for the $k$-th observation. In practice,
initially, we start with the prior distribution $\pi ({{\vec\theta
}})$ identified by expert opinions and external data only. Then, the
posterior distribution $\pi({ {\vec\theta }}\vert {\vec{x}})$ is
calculated, using (\ref{Chapter_combining_posterior}), when actual
data are observed. If there is a reason (for example, the new
control policy introduced in a bank), then this posterior
distribution can be adjusted by an expert and treated as the prior
distribution for subsequent observations.

\subsection{Estimating Prior}
\label{EstimatingPrior_sec} In general, the structural parameters of
the prior distributions can be estimated subjectively using expert
opinions (\emph{pure Bayesian approach}) or using data
(\emph{empirical Bayesian approach}). In a pure Bayesian approach,
the prior distribution is specified subjectively (that is, in the
context of OpRisk, using expert opinions). Berger \cite{Berger85}
lists several methods.

\begin{itemize}
\item \emph{Histogram approach}\index{histogram approach}: split the space of the parameter $\vec{\theta}$ into intervals and
specify the subjective probability for each interval. From this, the
smooth density of the prior distribution can be determined.

\item \emph{Relative Likelihood Approach}: compare the intuitive
likelihoods of the different values of $\vec\theta$. Again, the
smooth density of prior distribution can be determined. It is
difficult to apply this method in the case of unbounded parameters.

\item \emph{CDF determinations}: subjectively construct the
distribution function for the prior and sketch a smooth curve.

\item \emph{Matching a Given Functional Form}: find the prior
distribution parameters assuming some functional form for the prior
distribution to match prior beliefs (on the moments, quantiles, etc)
as close as possible.

\end{itemize}

The use of a particular method is determined by a specific problem
and expert experience. Usually, if the expected values for the
quantiles (or mean) and their uncertainties are estimated by the
expert then it is possible to fit the priors.

Often, expert opinions are specified for some quantities such as
quantiles or other risk characteristics rather than for the
parameters directly. In this case it might be better to assume some
priors for these quantities that will imply a prior for the
parameters. In general, given model parameters
$\vec\theta=(\theta_1,\ldots,\theta_n)$, assume that there are risk
characteristics $d_i=g_i(\vec\theta)$, $i=1,2,\ldots, n$ that are
well understood by experts. These could be some quantiles, expected
values, expected durations between losses exceeding high thresholds,
etc. Now, if experts specify the joint prior $\pi(d_1,\ldots,d_n)$,
then using transformation method the prior for
$\theta_1,\ldots,\theta_n$ is
\begin{equation}
\label{Prior_Transform_eq}
\pi(\vec\theta)=\pi(g_1(\vec\theta),\ldots,g_n(\vec\theta))\left|\frac{\partial
\left(g_1(\vec\theta),\ldots,g_n(\vec\theta)\right)}{\partial
\left(\theta_1,\ldots,\theta_n\right)}\right|,
\end{equation}
where $\left|\partial
\left(g_1(\vec\theta),\ldots,g_n(\vec\theta)\right)/\partial
\left(\theta_1,\ldots,\theta_n\right)\right|$ is the Jacobian
determinant of the transformation. Essentially, the main difficulty
in specifying a joint prior is due to a possible dependence between
the parameters. It is convenient to choose the characteristics (for
specification of the prior) such that independence can be assumed.
For example, if the prior for the quantiles $q_1,\ldots,q_n$
(corresponding to probability levels $p_1<p_2<\cdots<p_n$) is to be
specified, then to account for the ordering it might be better to
consider the differences
$$d_1=q_1, d_2=q_2-q_1, \ldots, d_n=q_n-q_{n-1}.$$
Then, it is reasonable to assume independence between these
differences and impose constraints $d_i>0$, $i=2,\ldots,n$. If
experts specify the marginal priors $\pi(d_1), \pi(d_2),\ldots,
\pi(d_n)$ (e.g. gamma priors) then the full joint prior is
$$
\pi(d_1,\ldots,d_n)=\pi(d_1)\times\pi(d_2)\times\cdots\times\pi(d_n)
$$
and the prior for parameters $\vec\theta$ is calculated by
transformation using (\ref{Prior_Transform_eq}). To specify the
$i$-th prior $\pi(d_i)$, an expert may use the approaches listed
above. For example, if $\pi(d_i)$ is $Gamma(\alpha_i,\beta_i)$, then
the expert may provide the mean and variational coefficient for
$\pi(d_i)$ (or median and 0.95 quantile) that should be enough to
determine $\alpha_i$ and $\beta_i$.

\begin{figure}[!htbp]
\centerline{
\includegraphics[scale=0.9]{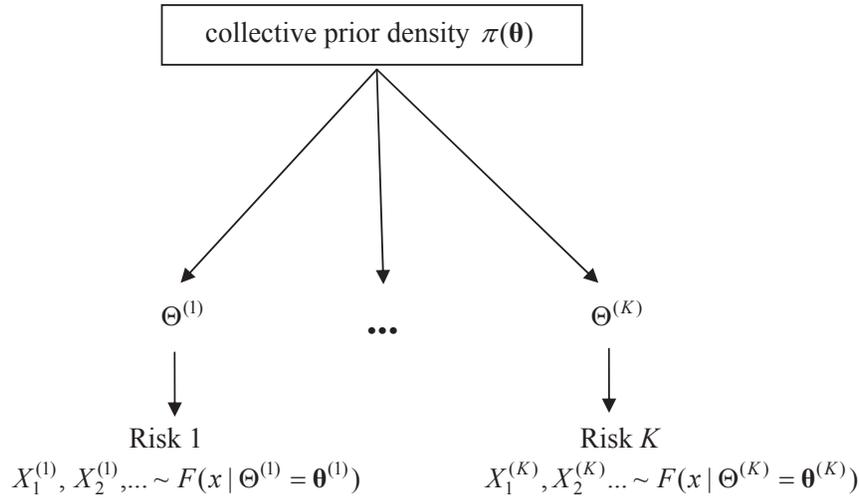}}
\caption{\small{Empirical Bayes approach -- interpretation of the
prior density $\pi(\vec{\theta})$. Here, $\vec{\Theta}^{(i)}$ is the
risk profile of the $i$-th risk. Given
$\vec{\Theta}^{(i)}=\vec{\theta}^{(i)}$, the risk data
$X_1^{(i)},X_2^{(i)},\ldots$ are generated from the distribution
$F(x|\vec{\theta}^{(i)})$. The risks are different having different
risk profiles $\vec{\theta}^{(i)}$, but
$\vec{\Theta}^{(1)},\ldots,\vec{\Theta}^{(K)}$ are distributed from
the same common density $\pi(\vec{\theta})$.}}
\label{Chapter_combining_TwoUrnModel_fig}
\end{figure}

Under empirical Bayesian approach, the parameter $\vec{\theta}$ is
treated as a random sample from the prior distribution. Then using
collective data of \emph{similar} risks, the parameters of the prior
are estimated using a marginal distribution of observations.
Depending on the model setup, the data can be collective industry
data, collective data in the bank, etc. To explain, consider $K$
similar risks where each risk has own risk profile
$\vec{\Theta}^{(i)}$, $i=1,\ldots,K$; see Figure
\ref{Chapter_combining_TwoUrnModel_fig}. Given
$\vec{\Theta}^{(i)}=\vec{\theta}^{(i)}$, the risk data
$X_1^{(i)},X_2^{(i)},\ldots$ are generated from the distribution
$F(x|\vec{\theta}^{(i)})$. The risks are different having different
risk profiles $\vec{\theta}^{(i)}$, but what they have in common is
that $\vec{\Theta}^{(1)},\ldots,\vec{\Theta}^{(K)}$ are distributed
from the same density $\pi(\vec{\theta})$. Then, one can find the
unconditional distribution of the data $\vec X$ and fit the prior
distribution using all data (across all similar risks). This could
be done, for example, by the maximum likelihood method or the method
of moments or even empirically. Consider, for example, $J$
\emph{similar} risk cells with the data $\{X_{k}^{(j)}$,
$k=1,2,\ldots\;$, $j = 1,\ldots,J\}$. This can be, for example, a
specific business line/event type risk cell in $J$ banks. Denote the
data over past years as
$\vec{X}^{(j)}=(X_{1}^{(j)},\ldots,X_{K_j}^{(j)})^\prime$, that is,
$K_j $ is the number of observations in bank $j$ over past years.
Assume that $X_{1}^{(j)} ,\ldots,X_{K_j}^{(j)}$ are conditionally
independent and identically distributed from the density $f(\cdot
\vert {\vec{\theta }}^j )$, for given ${\vec{\Theta
}}^{(j)}={\vec{\theta }}^{(j)}$.
 That is, the risk cells have different risk profiles ${\vec{\Theta }}^j $.
 Assume now that the risks are similar, in a sense that ${\vec{\Theta }}^{(1)},\ldots,{\vec{\Theta }}^{(J)}$ are
 independent and identically distributed from the same density $\pi
({\vec{\theta }})$. That is, it is assumed that the risk cells are
the same a priori (before we have any observations); see Figure
\ref{Chapter_combining_TwoUrnModel_fig}. Then the joint density of
all observations can be written as
\begin{equation}
\label{Chapter_combining_TotalIndustryLikelihood}
f(\vec{x}^{(1)},\ldots,\vec{x}^{(J)}) = \prod\limits_{j = 1}^J {\int
{\left[ {\prod\limits_{k = 1}^{K_j } {f(x_{k}^{(j)} \vert
{\vec{\theta }}^{(j)} )} } \right]\pi ({\vec{\theta }}^{(j)}
)d{\vec{\theta }}^{(j)} } } .
\end{equation}
The parameters of $\pi ({\vec{\theta }})$ can be estimated using the
maximum likelihood method by maximising
(\ref{Chapter_combining_TotalIndustryLikelihood}). The distribution
$\pi ({\vec{\theta }})$ is a prior distribution for the $j$-th cell.
Using internal data of the $j$-th risk cell, its posterior density
is calculated from (\ref{Chapter_combining_posterior}) as
\begin{equation} \pi({\vec{\theta }}^{(j)} \vert \vec{x}^{(j)}) =
\prod\limits_{k = 1}^{K_j } {f(x_{k}^{(j)} \vert {\vec{\theta
}}^{(j)} )} \pi ({\vec{\theta}}^{(j)} ),
\end{equation}
\noindent where $\pi ({\vec{\theta }})$ was fitted with MLE using
(\ref{Chapter_combining_TotalIndustryLikelihood}). The basic idea
here is that the estimates based on observations from all banks are
better then those obtained using smaller number of observations
available in the risk cell of a particular bank.

%\newpage

\subsection{Poisson Frequency}
\label{Chapter_BasicConcepts_PoissonGamma_sec} Consider the annual
number of events for a risk in one bank in year $t$ modelled as a
random variable from the Poisson distribution
$Poisson\left(\lambda\right)$. The intensity parameter $\lambda$ is
not known and the Bayesian approach models it as a random variable
$\Lambda$. Then the following model for years $t=1,2,\ldots,T,T+1$
(where $T+1$ corresponds to the next year) can be considered.

\begin{model_assumptions}
\label{Chapter_combining_PoissonGamma_assumptions}
 ~
\begin{itemize}

\item Suppose that, given {$\Lambda=\lambda $},
the data $N_1 ,\ldots,N_{T+1}$ are independent random variables from
the Poisson distribution, $Poisson(\lambda )$:

\begin{equation}
\Pr[N_t=n|\lambda] = e^{ - \lambda }\frac{\lambda ^n}{n!},\quad
\lambda \ge 0.
\end{equation}

\item The prior distribution for $\Lambda $ is a gamma distribution,
$Gamma(\alpha ,\beta )$, with a density

\begin{equation}
\pi (\lambda) = \frac{(\lambda / \beta )^{\alpha -
1}}{\mathrm{\Gamma} (\alpha )\beta }\exp ( - \lambda / \beta ),\quad
\lambda
> 0,\;\alpha > 0, \beta > 0.
\end{equation}

\end{itemize}
\end{model_assumptions}

\noindent That is, $\lambda $ plays the role of $\vec{\theta}$ and
$\vec{N}=(N_1 ,\ldots,N_{T})^\prime$ the role of $\vec{X}$ in
(\ref{Chapter_combining_posterior}).

~

\noindent\textbf{Posterior.} Given $\Lambda=\lambda $, under the
Model Assumptions \ref{Chapter_combining_PoissonGamma_assumptions},
$N_1 ,\ldots,N_T $ are independent and their  joint density, at
$\vec{N}=\vec{n}$, is given by

\begin{equation}
h({\vec{n}}\vert \lambda ) = \prod\limits_{i = 1}^T {e^{ - \lambda
}\frac{\lambda ^{n_i }}{n_i !}} .
\end{equation}

\noindent Thus, using formula (\ref{Chapter_combining_posterior}),
the posterior density is

\begin{equation}
\pi(\lambda \vert {\vec{n}}) \propto \frac{(\lambda / \beta
)^{\alpha - 1}}{\mathrm{\Gamma} (\alpha )\beta }\exp ( - \lambda /
\beta )\prod\limits_{i = 1}^T {e^{ - \lambda }\frac{\lambda ^{n_i
}}{n_i !}} \propto \lambda ^{{\alpha_T} - 1}\exp ( - \lambda /
{\beta_T}),
\end{equation}

\noindent which is $Gamma(\alpha_T,\beta_T)$, i.e. the same as the
prior distribution with updated parameters $\alpha_T$ and $\beta_T$
given by:
\begin{equation}
\label{Chapter_combining_PGposteriorParam}
 \alpha \to \alpha_T = \alpha + \sum\limits_{i = 1}^T {n_i } , \quad
 \beta \to \beta_T = \frac{\beta}{1 + \beta \times T}.
\end{equation}

~

\noindent\textbf{Improper constant prior.} It is easy to see that,
if the prior is constant (improper prior), i.e. $\pi(\lambda \vert
{\vec{n}}) \propto h({\vec{n}}\vert \lambda )$, then the posterior
is $Gamma(\alpha_T,\beta_T)$ with
\begin{equation}
\label{Chapter_combining_PGposteriorParam_improperConstPrior}
\alpha_T = 1 + \sum\limits_{i = 1}^T {n_i } , \quad \beta_T =
\frac{1}{T}.
\end{equation}
\noindent In this case, the mode of the posterior $\pi(\lambda| \vec
n)$ is
$\widehat{\lambda}_T^{\mbox{\tiny{MAP}}}=(\alpha_T-1)\beta_T=\frac{1}{T}\sum_{i=1}^{T}n_i$,
 which is the same as the maximum likelihood estimate (MLE)
$\widehat{\lambda}_T^{\mbox{\tiny{MLE}}}$ of $\lambda$.

 ~

\noindent\textbf{Predictive distribution.} Given data, the full
predictive distribution for $N_{T+1}$ is negative
binomial\index{negative binomial distribution},
$NegBin(\alpha_T,1/(1+\beta_T))$:
\begin{eqnarray}
\label{Chapter_BasicConcepts_PoissonGammaToNegBin_eq}
\Pr[N_{T+1}=m|\vec N=\vec{n}]&=&\int f(m|\lambda)\pi(\lambda|\vec n)d\lambda \nonumber\\
           &=&\int
           e^{-\lambda}\frac{\lambda^m}{m!}\frac{\lambda^{\alpha_T-1}}{(\beta_T)^{\alpha_T}
           \mathrm{\Gamma}(\alpha_T)}e^{-\lambda/\beta_T}d\lambda\nonumber\\
            &=&\frac{(\beta_T)^{-\alpha_T}}{\mathrm{\Gamma}(\alpha_T)m!}\int
            e^{-(1+1/\beta_T)\lambda}\lambda^{\alpha_T+m-1}d\lambda
            \nonumber\\
            &=&\frac{\mathrm{\Gamma}(\alpha_T+m)}{\mathrm{\Gamma}(\alpha_T)m!}\left(\frac{1}{1+\beta_T}\right)^{\alpha_T}
            \left(\frac{\beta_T}{1+\beta_T}\right)^m.
\end{eqnarray}
It is assumed that given $\Lambda=\lambda$, $N_{T+1}$ and $\vec N$
are independent. The expected number of events over the next year,
given past observations, $\mathrm{E}[N_{T + 1} \vert {\vec{N}}]$,
i.e. mean of $NegBin(\alpha_T,1/(1+\beta_T))$ (which is also a mean
of the posterior distribution in this case), allows for a good
interpretation as follows:
\begin{eqnarray}
\label{Chapter_combining_PoissonFreq_Expected_eq} \mathrm{E}[N_{T +
1} \vert {\vec{N}}=\vec{n}] = \mathrm{E}[\lambda \vert
{\vec{N}}=\vec{n}] = \alpha_T \beta_T &=& \beta \frac{\alpha +
\sum_{i = 1}^T {n_i } }{1 +
\beta \times T} \nonumber\\
&=& w_T \widehat{\lambda}_T^{\mbox{\tiny{MLE}}} + (1 -
w_T)\lambda_0.
\end{eqnarray}
\noindent Here,
\begin{itemize}

\item $\widehat{\lambda}_T^{\mbox{\tiny{MLE}}} = \frac{1}{T}\sum_{i = 1}^T {n_i }$ is the
estimate of $\lambda $ using the observed counts only;

\item $\lambda _0 = \alpha \beta $ is the estimate of $\lambda $
using a prior distribution only (e.g. specified by expert);

\item $w_T = \frac{ T\beta}{ T\beta + 1}$ is the credibility weight in [0,1)\index{weight!credibility}
used to combine $\lambda _0 $ and
$\widehat{\lambda}_T^{\mbox{\tiny{MLE}}}$.

\end{itemize}

\begin{remark}
~
\begin{itemize}
\item As the number of observed years $T$ increases, the credibility weight $w_T$
increases and vice versa. That is, the more observations we have,
the greater credibility weight we assign to the estimator based on
the observed counts, while the lesser credibility weight is attached
to the expert opinion estimate. Also, the larger the volatility of
the expert opinion (larger $\beta )$, the greater credibility weight
is assigned to observations.

\item Recursive calculation of the posterior distribution is very
simple. That is, consider observed annual counts $n_1 ,n_2
,\ldots,n_k ,\ldots$ , where $n_k $ is the number of events in the
$k$-th year. Assume that the prior $Gamma(\alpha ,\beta )$ is
specified initially, then the posterior $\pi (\lambda \vert n_1
,\ldots,n_k )$ after the $k$-th year is a gamma distribution,
$Gamma(\alpha_k ,\beta_k )$, with $\alpha_k = \alpha + \sum_{i =
1}^k {n_i } $ and $\beta_k = \beta / (1 + \beta \times k)$. Observe
that,
\begin{equation}
\label{Chapter_combining_PGiterativePosterEst} \alpha_k = \alpha_{k
- 1} + n_k,\quad \beta_k = \frac{\beta _{k - 1}}{1 + \beta_{k - 1}
}.
\end{equation}
This leads to a very efficient recursive scheme, where the
calculation of posterior distribution parameters is based on the
most recent observation and parameters of posterior distribution
calculated just before this observation.
\end{itemize}

\end{remark}

~

\noindent\textbf{Estimating prior.} Suppose that the annual
frequency of the OpRisk losses $N$ is modelled by the Poisson
distribution, $Poisson(\Lambda=\lambda )$, and the prior density
$\pi (\lambda)$ for $\Lambda$ is $Gamma(\alpha ,\beta )$. Then,
$\mathrm{E}[N\vert \Lambda ] = \Lambda $ and $\mathrm{E}[\Lambda ] =
\alpha \times \beta $. The expert may estimate the expected number
of events but cannot be certain in the estimate. One could say that
the expert's ``best'' estimate for the expected number of events
corresponds to $\mathrm{E}[\mathrm{E}[N\vert \Lambda ]] =
\mathrm{E}[\Lambda ]$. If the expert specifies $\mathrm{E}[\Lambda
]$ and an uncertainty that the ``true'' $\lambda$ for next year is
within the interval [$a$,$b$] with a probability $\Pr [a \le \Lambda
\le b] = p$ (it may be convenient to set $p = 2 / 3)$, then the
equations

\begin{equation}
\label{Chapter_combining_PoissonGammaExpert}
\begin{array}{l}
 \mathrm{E}[\Lambda ] = \alpha \times \beta , \\
 \Pr [a \le \Lambda \le b] = p = \int\limits_a^b {\pi (\lambda \vert \alpha
,\beta )d\lambda } = F_{\alpha ,\beta }^{(G)}(b) - F_{\alpha ,\beta
}^{(G)}
(a) \\
 \end{array}
\end{equation}

\noindent can be solved numerically to estimate the structural
parameters $\alpha $ and $\beta $. Here, $F_{\alpha ,\beta }^{(G)}
(\cdot)$ is the gamma distribution, $Gamma(\alpha,\beta)$, i.e.

\begin{equation*}
F_{\alpha ,\beta }^{(G)} [y] = \int\limits_0^y {\frac{x^{\alpha -
1}}{\mathrm{\Gamma} (\alpha )\beta ^\alpha }\exp \left( { -
\frac{x}{\beta }} \right)dx} .
\end{equation*}

In the insurance industry, the uncertainty for the ``true'' $\lambda
$ is often measured in terms of the coefficient of variation,
$\mathrm{Vco}[\Lambda ] = \sqrt {\mathrm{Var}[\Lambda]} /
\mathrm{E}[\Lambda ]$. Given the expert estimates for
$\mathrm{E}[\Lambda ] = \alpha \beta $ and $\mathrm{Vco}[\Lambda ] =
1 / \sqrt \alpha ,$ the structural parameters $\alpha $ and $\beta $
are easily estimated.

\subsection{Numerical example}\label{Chapter_combining_PoissonGamma_example} If the
expert specifies $\mathrm{E}[\Lambda ] = 0.5$ and $\Pr [0.25 \le
\Lambda \le 0.75] = 2 / 3,$ then we can fit a prior distribution
$Gamma(\alpha \approx 3.407,\;\beta \approx 0.147)$ by solving
(\ref{Chapter_combining_PoissonGammaExpert}). Assume now that the
bank experienced no losses over the first year (after the prior
distribution was estimated). Then, using formulas
(\ref{Chapter_combining_PGiterativePosterEst}), the posterior
distribution parameters are $\widehat {\alpha }_1 \approx 3.407 + 0
= 3.407,$ $\widehat {\beta }_1 \approx 0.147 / (1 + 0.147) \approx
0.128$ and the estimated arrival rate using the posterior
distribution is $\widehat {\lambda }_1 = \widehat {\alpha }_1 \times
\widehat {\beta }_1 \approx 0.436.$ If during the next year no
losses are observed again, then the posterior distribution
parameters are $\widehat {\alpha }_2 = \widehat {\alpha }_1 + 0
\approx 3.407,$ $\widehat {\beta }_2 = \widehat {\beta }_1 / (1 +
\widehat {\beta }_1 ) \approx 0.113$ and $\widehat {\lambda }_2 =
\widehat {\alpha }_2 \times \widehat {\beta }_2 \approx 0.385.$
Subsequent observations will update the arrival rate estimator
correspondingly using formulas
(\ref{Chapter_combining_PGiterativePosterEst}). Thus, starting from
the expert specified prior, observations regularly update (refine)
the posterior distribution. The expert might reassess the posterior
distribution at any point in time (the posterior distribution can be
treated as a prior distribution for the next period), if new
practices/policies were introduced in the bank that affect the
frequency of the loss. That is, if we have a new policy at time $k$,
the expert may reassess the parameters and replace $\widehat {\alpha
}_k $ and $\widehat {\beta }_k $ by $\widehat {\alpha }_k^\ast $ and
$\widehat {\beta }_k^\ast $ respectively.

\begin{figure}[!htbp]
\begin{center}
\centerline{
\includegraphics[scale=0.8]{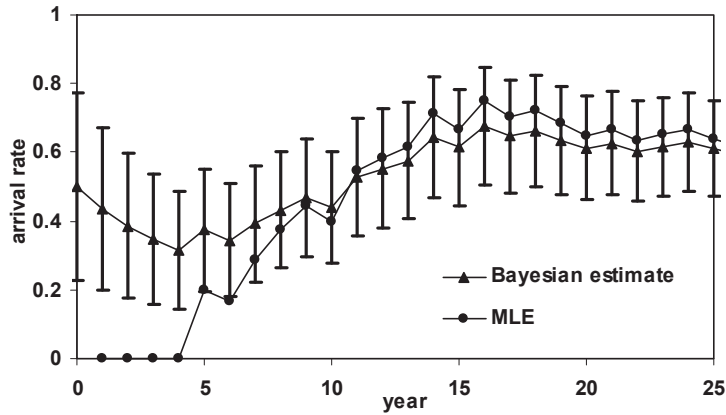}}
\caption{\small{The Bayesian and the standard maximum likelihood
estimates of the arrival rate vs the observation year; see Section
\ref{Chapter_combining_PoissonGamma_example} for details.}}
\label{Chapter_combining_FreqEst_fig1}
\end{center}
\end{figure}

In Figure \ref{Chapter_combining_FreqEst_fig1}, we show the
posterior best estimate for the arrival rate $\widehat {\lambda }_k
= \widehat {\alpha }_k \times \widehat {\beta }_k $, $k =
1,\ldots,15$ (with the prior distribution as in the above example),
when the annual number of events $N_k $, $k = 1,\ldots,25$ are
simulated from $Poisson(\lambda = 0.6)$ and the realized samples for
25 years are
$n_{1:25}=(0,0,0,0,1,0,1,1,1,0,2,1,1,2,0,2,0,1,0,0,1,0,1,1,0).$

On the same figure, we show the standard maximum likelihood estimate
of the arrival rate $\widehat{\lambda}_k^{\mbox{\tiny{MLE}}} =
\textstyle{1 \over k}\sum\nolimits_{i = 1}^k {n_i } $. After
approximately 8 years, the estimators are very close to each other.
However, for a small number of observed years, the Bayesian estimate
is more accurate as it takes the prior information into account.
Only after 12 years do both estimators converge to the true value of
0.6 (this is because the bank was very lucky to have no events
during the first four years). Note that for this example we assumed
the prior distribution with a mean equal to 0.5, which is different
from the true arrival rate. Thus this example shows that an
initially incorrect prior estimator is corrected by the observations
as they become available. It is interesting to observe that, in year
14, the estimators become slightly different again. This is because
the bank was unlucky to experience event counts (1, 1, 2) in the
years (12, 13, 14). As a result, the maximum likelihood estimate
becomes higher than the true value, while the Bayesian estimate is
more stable (smooth) with respect to the unlucky years. If this
example is repeated with different sequences of random numbers, then
one would observe quite different maximum likelihood estimates (for
small $k)$ and more stable Bayesian estimates.

Finally we note that the standard deviation of the posterior
distribution $Gamma(\alpha_k,\beta_k)$ is large for small $k$. It is
indicated by the error bars in Figure
\ref{Chapter_combining_FreqEst_fig1} and calculated as
$\beta_k\sqrt{\alpha_k}$.

\subsection{The Lognormal $\mathcal{LN}(\mu,\sigma)$ Severity}
\label{Chapter_coimbining_LognormalUnknownMu_sec} Assume that the
loss severity for a risk in one bank is modelled as a random
variable from a lognormal distribution, $\mathcal{LN}(\mu ,\sigma)$,
whose density is
\begin{equation}
\label{Chapter_combining_LNpdf} f(x\vert \mu ,\sigma ) =
\frac{1}{x\sqrt {2\pi \sigma ^2} }\exp \left( { - \frac{(\ln x - \mu
)^2}{2\sigma ^2}} \right).
\end{equation}
This distribution often gives a good fit for operational loss data.
Also, it belongs to a class of heavy-tailed (subexponential)
distributions. The parameters $\mu$ and $\sigma$ are not known and
the Bayesian approach models these as a random variables
$\Theta_\mu$ and $\Theta_\sigma$ respectively. We assume that the
losses over the years $t=1,2,\ldots,T$ are observed and should be
modelled for next year $T+1$. To simplify notation, we denote the
losses over past $T$ years as $X_1 ,\ldots,X_n$ and the future
losses are $X_{n+1},\ldots\;$. Then the model can be structured as
follows. For simplicity, assume that $\sigma$ is known and $\mu$ is
unknown. The case where both $\sigma$ and $\mu$ are unknown can be
found in Shevchenko \cite[section 4.3.5]{shevchenko2011}.

\begin{model_assumptions}
\label{Chapter_combining_Lognormal_normal_assumptions} ~
\begin{itemize}

\item Suppose that, given $\sigma$ and $\Theta_\mu=\mu$,
the data $X_1 ,\ldots,X_n,\ldots$ are independent random variables
from $\mathcal{LN}(\mu ,\sigma )$. That is, $Y_i = \ln X_i $, $i =
1,2,\ldots$ are distributed from the normal distribution
$\mathcal{N}(\mu ,\sigma )$.

\item Assume that parameter $\sigma$ is known and the prior
distribution for $\Theta_\mu $ is the normal distribution,
$\mathcal{N}(\mu _0 ,\sigma _0 )$. That is the prior density is
\begin{equation}
\label{Chapter_combining_NormalPDF} \pi (\mu) = \frac{1}{\sigma _0
\sqrt {2\pi } }\exp \left( { - \frac{(\mu - \mu _0 )^2}{2\sigma _0^2
}} \right).
\end{equation}

\end{itemize}

\noindent Denote the losses over past years as ${\vec{X}} = (X_1
,\ldots,X_n )^\prime$ and corresponding log-losses as ${\vec{Y}} =
(Y_1 ,\ldots,Y_n )^\prime$. Note that $\mu $ plays the role of
${\vec{\theta }}$ in (\ref{Chapter_combining_posterior}).
\end{model_assumptions}

~

\noindent\textbf{Posterior.} Under the above assumptions, the joint
density of the data over past years (conditional on $\sigma$ and
$\Theta_\mu=\mu$) at position $\vec{Y}=\vec{y}$ is

\begin{equation}
h({\vec{y}}\vert \mu ,\sigma ) = \prod\limits_{i = 1}^n
{\frac{1}{\sigma \sqrt {2\pi } }\exp \left( { - \frac{(y_i - \mu
)^2}{2\sigma ^2}} \right)} .
\end{equation}

\noindent Then, using formula (\ref{Chapter_combining_posterior}),
the posterior density can be written as

\begin{eqnarray}
\label{Chapter_combining_LNposterior} \pi(\mu \vert {\vec {y}})
&\propto& \frac{\exp \left( { - \frac{(\mu - \mu _0 )^2}{2\sigma
_0^2 }} \right)}{\sigma _0 \sqrt {2\pi } }\prod\limits_{i = 1}^n
{\frac{\exp \left( { - \frac{(y_i - \mu )^2}{2\sigma ^2}}
\right)}{\sigma \sqrt {2\pi } }} \nonumber\\
&\propto& \exp \left( { - \frac{(\mu - {\mu }_{0,n})^2}{2{\sigma
}_{0,n}^2 }} \right),
\end{eqnarray}

\noindent that corresponds to a normal distribution,
$\mathcal{N}({\mu }_{0,n} ,{\sigma }_{0,n} )$, i.e. the same as the
prior distribution with updated parameters
\begin{eqnarray}
\label{Chapter_combining_LNposterCalc}
 \mu_0 \to {\mu }_{0,n} &=& \frac{\mu _0 + \omega \sum\limits_{i = 1}^n {y_i }}
{1 + n\times \omega}, \\
 \sigma _0^2 \to {\sigma }_{0,n}^2 &=& \frac{\sigma_0^2}{1 + n\times \omega
},\quad \mbox{where}\;\;\omega = \sigma _0^2 / \sigma ^2.
\end{eqnarray}

~

The expected value of $Y_{n + 1} $ (given past observations),
$\mathrm{E}[Y_{n + 1} \vert {\vec{Y}}=\vec{y}]$, allows for a good
interpretation, as follows:
\begin{eqnarray}
\label{Chapter_combining_LNpredmean} \mathrm{E}[Y_{n + 1} \vert
{\vec{Y}}=\vec{y}] = \mathrm{E}[\Theta_\mu \vert {\vec{Y}}=\vec{y}]
= {\mu }_{0,n} &=& \frac{\mu_0 + \omega \sum\limits_{i = 1}^n {y_i }
}{1 + n\times
\omega }\nonumber\\
& = &w_n\overline{y}_n + (1 - w_n)\mu _0 ,
\end{eqnarray}

\noindent where
\begin{itemize}

\item $\overline{y}_n = \frac{1}{n}\sum\limits_{i = 1}^n {y_i }$ is the
estimate of $\mu$ using the observed losses only;

\item $\mu_0$ is the estimate of $\mu $ using a prior distribution
only (e.g. specified by expert);

\item $w_n = \frac{n}{n + \sigma ^2 / \sigma _0^2}$ is the credibility
weight in [0,1)\index{weight!credibility} used to combine $\mu _0 $
and $\overline{y}_n$.

\end{itemize}

~

\begin{remark}
~
\begin{itemize}
\item As the number of observations increases, the credibility weight
$w$ increases and vice versa. That is, the more observations we have
the greater weight we assign to the estimator based on the observed
counts and the lesser weight is attached to the expert opinion
estimate. Also, larger uncertainty in the expert opinion $\sigma
_0^2 $ leads to a higher credibility weight for observations and
larger volatility of observations $\sigma ^2$ leads to a higher
credibility weight for expert opinions.

\item The posterior distribution can be calculated recursively as
follows.\textbf{ }Consider the data $Y_1 ,Y_2 ,\ldots,Y_k ,\ldots$ .
Assume that the prior distribution, $\mathcal{N}(\mu _0 ,\sigma_0)$,
is specified initially, then the posterior density ${\pi }(\mu \vert
y_1 ,\ldots,y_k )$ after the $k$-th event is the normal distribution
$\mathcal{N}({\mu }_{0,k}
 ,{\sigma }_{0,k} )$ with
$${\mu }_{0,k} = \frac{\mu _0 + \omega \sum\limits_{i = 1}^k {y_i }
}{1 + k\times \omega},\quad{\sigma }_{0,k}^2  = \frac{\sigma _0^2}{1
+ k\times \omega},$$ \noindent where $\omega = \sigma _0^2 / \sigma
^2$. It is easy to show that
\begin{equation}
\label{Chapter_combining_LNrecursivePosterCalc} {\mu }_{0,k} =
\frac{{\mu }_{0,k - 1} + \omega_{k-1} y_k }{1 +
\omega_{k-1}},\;\quad {\sigma }_{0,k}^2 = \frac{\sigma ^2
\omega_{k-1} }{1 + \omega_{k-1}}
\end{equation}
\noindent with $\omega_{k-1}={\sigma }_{0,k-1}^2 / \sigma ^2$. That
is, calculation of the posterior distribution parameters can be
based on the most recent observation and the parameters of the
posterior distribution calculated just before this observation.

\item Estimation of prior for the parameters of lognormal distribution is considered in Shevchenko and W\"{u}thrich \cite{ShWu06}.
\end{itemize}
\end{remark}

\section{Bayesian Method to Combine Three Data Sources}
\label{Chapter_combining_3sources_section} In the previous section
we
 showed how to combine two data sources: expert opinions and
internal data; or external data and internal data. In order to
estimate the risk capital of a bank and to fulfill the Basel II
requirements, risk managers have to take into account internal data,
relevant external data (industry data) and expert opinions. The aim
of this section is to provide an example of methodology to be used
to combine these three sources of information. Here, we follow the
approach suggested in Lambrigger et al \cite{LaShWu07}. As in the
previous section, we consider one risk cell only. In terms of
methodology we go through the following steps:

\begin{itemize}
\item In any risk cell, we model the loss frequency and the loss
severity by parametric distributions (e.g. Poisson for the frequency
or Pareto, lognormal, etc. for the severity). For the considered
bank, the unknown parameter vector $\vec{\theta}$ (for example, the
Poisson parameter or the Pareto tail index) of these distributions
has to be quantified.

\item A priori, before we have any company specific information, only
industry data are available. Hence, the best prediction of our bank
specific parameter $\vec{\theta}$ is given by the belief in the
available external knowledge such as the provided industry data.
This unknown parameter of interest is modelled by a prior
distribution (structural distribution) corresponding to a random
vector $\vec{\Theta}$. The parameters of the prior distribution
(hyper-parameters) are estimated using data from the whole industry
by, for example, maximum likelihood estimation. If no industry data
are available, the prior distribution could come from a ``super
expert'' that has an overview over all banks.

\item The true bank specific parameter
$\vec{\theta}_0$ is treated as a realisation of $\vec{\Theta}$. The
prior distribution of a random vector $\vec{\Theta}$ corresponds to
the whole banking industry sector, whereas $\vec{\theta}$ stands for
the unknown underlying parameter set of the bank being considered.
Due to the variability amongst banks, it is natural to model
$\vec{\theta}$ by a probability distribution. Note that
$\vec{\Theta}$ is random with known distribution, whereas
$\vec{\theta}_0$ is deterministic but unknown.

\item As time passes, internal data $\vec{X}=(X_1,\ldots,X_K)^\prime$
as well as expert opinions
$\vec{\Delta}=(\Delta_{1},\ldots,\Delta_{M})^\prime$ about the
underlying parameter $\vec{\theta}$ become available. This affects
our belief in the distribution of $\vec{\Theta}$ coming from
external data only and adjust the prediction of $\vec{\theta}_0$.
The more information on $\vec{X}$ and $\vec{\Delta}$ we have, the
better we are able to predict $\vec{\theta}_0$. That is, we replace
the prior density $\pi(\vec{\theta})$ by a conditional density of
$\vec{\Theta}$ given $\vec{X}$ and $\vec{\Delta}$.

\end{itemize}

 In order to determine the posterior density $\pi(\vec{\theta} | \vec{x},
 \vec{\delta})$,
consider the joint conditional density of observations and expert
opinions (given the parameter vector $\vec{\theta}$):
\begin{equation}
\label{Chapter_combining_eq:independence}
h(\vec{x},\vec{\delta}|\vec{\theta}) = h_1(\vec{x}|\vec{\theta})
h_2(\vec{\delta}|\vec{\theta}),
\end{equation}
where $h_1$ and $h_2$ are the conditional densities (given
$\vec{\Theta}=\vec{\theta}$) of $\vec{X}$ and $\vec{\Delta}$,
respectively. Thus $\vec{X}$ and $\vec{\Delta}$ are assumed to be
conditionally independent given $\vec{\Theta}$.

\begin{remark}
~
\begin{itemize}
\item Notice that, in this way, we naturally combine external data information,
$\pi(\vec{\theta})$, with internal data $\vec{X}$ and expert opinion
$\vec{\Delta}$.
\item In classical Bayesian inference (as it is used, for example, in actuarial science),
one usually combines only two sources of information as described in
the previous sections. Here, we combine three sources simultaneously
using an appropriate structure, that is, equation
(\ref{Chapter_combining_eq:independence}).
\item Equation (\ref{Chapter_combining_eq:independence}) is quite a reasonable
assumption. Assume that the true bank specific parameter is
$\vec{\theta}_0$. Then, (\ref{Chapter_combining_eq:independence})
says that the experts in this bank estimate $\vec{\theta}_0$ (by
their opinion $\vec{\Delta}$) independently of the internal
observations. This makes sense if the experts specify their opinions
regardless of the data observed. Otherwise we should work with the
joint distribution $h(\vec{x},\vec{\delta}|\vec{\theta})$.
\end{itemize}

\end{remark}
We further assume that observations as well as expert opinions are
conditionally independent and identically distributed, given
$\vec{\Theta}=\vec{\theta}$, so that
\begin{eqnarray}
h_1(\vec{x}|\vec{\theta}) &=& \prod_{k=1}^{K} f_1(x_k | \vec{\theta}),\\
h_2(\vec{\delta}|\vec{\theta}) &=& \prod_{m=1}^{M} f_2(\delta_{m} |
\vec{\theta}),
\end{eqnarray}
where $f_1$ and $f_2$ are the marginal densities of a single
observation and a single expert opinion, respectively. We have
assumed that all expert opinions are identically distributed, but
this can be generalised easily to expert opinions having different
distributions.

Here, the unconditional parameter density $\pi(\vec{\theta})$ is the
\emph{prior} density, whereas the conditional parameter density
$\pi(\vec{\theta}| \vec{x},\vec{\delta})$ is the \emph{posterior}
density. Let $h(\vec{x},\vec{\delta})$ denote the unconditional
joint density of the data $\vec{X}$ and expert opinions
$\vec{\Delta}$. Then, it follows from Bayes's theorem that
\begin{equation}
 h(\vec{x},\vec{\delta}|\vec{\theta}) \pi(\vec{\theta}) =
\pi(\vec{\theta}| \vec{x},\vec{\delta}) h(\vec{x},\vec{\delta}).
\end{equation} Note that the unconditional density
$h(\vec{x},\vec{\delta})$ does not depend on $\vec{\theta}$ and thus
the posterior density is given by
\begin{equation}
\label{Chapter_combining_aposterioridistribution} \pi(\vec{\theta}|
\vec{x},\vec{\delta}) \propto \pi(\vec{\theta}) \prod_{k=1}^{K}
f_1(x_k | \vec{\theta}) \prod_{m=1}^{M} f_2(\delta_{m} |
\vec{\theta}).
\end{equation}
\noindent For the purposes of OpRisk, it should be used to estimate
the predictive distribution of future losses.

\subsection{Modelling Frequency: Poisson Model}
\label{Chapter_combining_SectionLossFreq3dataSources} To model the
loss frequency for OpRisk in a risk cell, consider the following
model.

\begin{model_assumptions}[Poisson-gamma-gamma]
\label{Chapter_combining_model:PoissonGammaGamma} Assume that a risk
cell in a bank has a scaling factor $V$ for the frequency in a
specified risk cell (it can be the product of several economic
factors such as the gross income, the number of transactions or the
number of staff).
\begin{itemize}
\item[a)] Let $\Lambda \sim Gamma (\alpha_0, \beta_0)$ be a gamma distributed random variable with shape
parameter $\alpha_0>0$ and scale parameter $\beta_0>0$, which are
estimated from (external) market data. That is, the density of
$Gamma (\alpha_0, \beta_0)$, plays the role of $\pi(\vec{\theta})$
in (\ref{Chapter_combining_aposterioridistribution}).
\item[b)] Given $\Lambda=\lambda$, the annual frequencies, $N_{1}, \ldots, N_{T},N_{T+1}$, where $T+1$ refers to next year,
are assumed to be independent and identically distributed with
$N_{t}\sim Poisson(V \lambda)$. That is, $f_1(\cdot|\lambda)$ in
(\ref{Chapter_combining_aposterioridistribution}) corresponds to the
probability mass function of a $Poisson(V \lambda)$ distribution.
\item[c)] A financial company has $M$ expert opinions $\Delta_{m}$, $1 \leq m \leq M$,
about the intensity parameter $\Lambda$. Given $\Lambda =\lambda$,
$\Delta_{m}$ and $N_{t}$ are independent for all $t$ and $m$, and
$\Delta_{1},\ldots,\Delta_M$ are independent and identically
distributed with $\Delta_m \sim Gamma (\xi,\lambda/\xi)$, where
$\xi$ is a known parameter. That is, $f_2(\cdot|\lambda)$
corresponds to the density of a $Gamma(\xi,\lambda/\xi)$
distribution.
\end{itemize}
\end{model_assumptions}

\begin{remark}
~
\begin{itemize}
\item The parameters $\alpha_0$ and $\beta_0$ in
Model Assumptions \ref{Chapter_combining_model:PoissonGammaGamma}
are hyper-parameters\index{hyper-parameters} (parameters for
parameters) and can be estimated using the maximum likelihood method
or the method of moments.

\item In Model Assumptions \ref{Chapter_combining_model:PoissonGammaGamma} we assume
\begin{equation}
\mathrm{E}[\Delta_{m}|\Lambda]=\Lambda, \quad 1 \leq m \leq M,
\end{equation} that is, expert opinions are unbiased. A
possible bias might only be recognised by the regulator, as he alone
has the overview of the whole market.
\end{itemize}

\end{remark}
Note that the \emph{coefficient of variation} of the conditional
expert opinion $\Delta_{m}|\Lambda$ is
$$\mathrm{Vco}[\Delta_{m}|\Lambda]=
(\mathrm{Var}[\Delta_{m}|\Lambda)])^{1/2} /
\mathrm{E}[\Delta_{m}|\Lambda] = 1/\sqrt{\xi},$$ and thus is
independent of $\Lambda$. This means that $\xi$, which characterises
the uncertainty in the expert opinions, is independent of the true
bank specific $\Lambda$. For simplicity, we have assumed that all
experts have the same conditional coefficient of variation and thus
have the same credibility. Moreover, this allows for the estimation
of $\xi$ as
\begin{equation}
\label{Chapter_combining_eq:xiestimate} \widehat{\xi}=
(\widehat{\mu}/ \widehat{\sigma})^2,
\end{equation}
\noindent where
\begin{equation*}
 \widehat{\mu} =
\frac{1}{M} \sum_{m=1}^{M} \delta_{m} \quad \textnormal{and} \quad
\widehat{\sigma}^2 = \frac{1}{M - 1} \sum_{m=1}^{M} (\delta_{m} -
\widehat{\mu})^2, \quad M \geq 2.
\end{equation*}

In a more general framework the parameter $\xi$ can be estimated,
for example, by maximum likelihood.

In the insurance practice $\xi$ is often specified by the regulator
denoting a lower bound for expert opinion uncertainty; e.g. Swiss
Solvency Test, see Swiss Financial Market Supervisory Authority
(\cite{SST06}, appendix 8.4). If the credibility differs among the
experts, then $\mathrm{Vco}[\Delta_{m}|\Lambda]$ should be estimated
for all $m$, $1 \leq m \leq M$. Admittedly, this may often be a
challenging issue in practice.

\begin{remark}
This model can be extended to a model where one allows for more
flexibility in the expert opinions. For convenience, it is preferred
that experts are conditionally independent and identically
distributed, given $\Lambda$. This has the advantage that there is
only one parameter, $\xi$, that needs to be estimated.
\end{remark}

Using the notation from Section
\ref{Chapter_combining_3sources_section}, the posterior density of
$\Lambda$, given the losses up to year $K$ and the expert opinions
of $M$ experts, can be calculated. Denote the data over past years
as follows:
\begin{eqnarray*}
\vec{N} &=& (N_{1},\ldots,N_{T})^\prime,\\
\vec{\Delta}&=& (\Delta_1,\ldots,\Delta_M)^\prime.
\end{eqnarray*}
Also, denote the arithmetic means by
\begin{equation} \overline{N} = \frac{1}{T} \sum_{t=1}^{T}
N_{t}, \quad \overline{\Delta} = \frac{1}{M} \sum_{m=1}^{M}
\Delta_m, \quad \textnormal{etc.}
\end{equation} Then, the posterior density is given by the
following theorem.

\begin{theorem}
\label{Chapter_combining_GIGThm} Under Model Assumptions
\ref{Chapter_combining_model:PoissonGammaGamma}, given loss
information $\vec{N}=\vec{n}$ and expert opinion
$\vec{\Delta}=\vec{\delta}$, the posterior density of $\Lambda$ is
\begin{equation}
\label{Chapter_combining_GIG} \pi(\lambda | \vec{n}, \vec{\delta}) =
\frac{(\omega/\phi)^{(\nu+1)/2}}{2 K_{\nu+1}(2 \sqrt{\omega \phi})}
\lambda^\nu e^{-\lambda \omega - \lambda^{-1} \phi},
\end{equation} with
\begin{eqnarray}
\label{Chapter_combining_eq:nuomegaphi}
\nu &=& \alpha_0 - 1 - M \xi + T \overline{n}, \nonumber\\
\omega &=& V T + \frac{1}{\beta_0},\\
\phi &=& \xi M \overline{\delta},\nonumber
\end{eqnarray}
and
\begin{equation} \label{Chapter_combining_def:modifiedBessel3}
K_{\nu +1}(z) = \frac{1}{2} \int_0^{\infty} u^\nu
e^{-z(u+1/u)/2}\textnormal{d}u.
\end{equation}
\end{theorem}
Here, $K_{\nu}(z)$ is a modified Bessel function of the third kind;
see for instance Abramowitz and Stegun (\cite{AbSt65}, p. 375).

\begin{proof}
Model Assumptions \ref{Chapter_combining_model:PoissonGammaGamma}
applied to (\ref{Chapter_combining_aposterioridistribution}) yield
\begin{eqnarray}
{\pi}(\lambda | \vec{n}, \vec{\delta}) &\propto&
\lambda^{\alpha_0-1} e^{-\lambda/\beta_0} \prod_{t=1}^{T} e^{-V
\lambda} \frac{(V \lambda)^{n_{t}}}{n_{t}!}
\prod_{m=1}^{M} \frac{(\delta_m)^{\xi-1}}{(\lambda/\xi)^{\xi}} e^{-\delta_m\xi/\lambda} \nonumber\\
&\propto& \lambda^{\alpha_0-1} e^{-\lambda/\beta_0} \prod_{t=1}^{T}
e^{-V \lambda} \lambda^{n_{t}}
\prod_{m=1}^{M} (\xi/\lambda)^{\xi}  e^{-\delta_{m} \xi /\lambda} \nonumber\\
&\propto& \lambda^{\alpha_0-1-M \xi + T \overline{n}}
\exp{\left(-\lambda \left(V T + \frac{1}{\beta_0} \right)
-\frac{1}{\lambda} \xi M \overline{\delta} \right)}.\nonumber
\label{Chapter_combining_formula:numberposteriori}
\end{eqnarray}
\end{proof}
\begin{remark}
~
\begin{itemize}
\item A distribution with density (\ref{Chapter_combining_GIG}) is known as the generalised inverse Gaussian distribution
GIG$(\omega,\phi,\nu)$\index{GIG distribution}. This is a well-known
distribution with many applications in finance and risk management;
see McNeil et al \cite[p. 75 and p. 497]{McFrEm05}.

\item In comparison with the classical Poisson-gamma case of combining two sources of information
(considered in Section
\ref{Chapter_BasicConcepts_PoissonGamma_sec}), where the posterior
is a gamma distribution, the posterior $\pi(\lambda|\cdot)$ in
(\ref{Chapter_combining_formula:numberposteriori}) is more
complicated. In the exponent, it involves  both $\lambda$ and
$1/\lambda$. Note that expert opinions enter via the term
$1/\lambda$ only.
\item Observe that the classical exponential dispersion family with associated conjugates
(see Chapter 2.5 in B\"{u}hlmann and Gisler \cite{BuGi05}) allows
for a natural extension to GIG-like distributions. In this sense the
GIG distributions enlarge the classical Bayesian inference theory on
the exponential dispersion family.
\end{itemize}
\end{remark}

For our purposes it is interesting to observe how the posterior
density transforms when new data from a newly observed year arrive.
Let $\nu_k$, $\omega_k$ and $\phi_k$ denote the parameters for the
data $(N_{1},\ldots,N_{k})$ after $k$ accounting years.
Implementation of the update processes is then given by the
following equalities (assuming that expert opinions do not change).
\begin{eqnarray}
\nu_{k+1} &=& \nu_k + n_{k+1},\nonumber\\
\omega_{k+1} &=& \omega_k + V,\\
\phi_{k+1} &=& \phi_k.\nonumber
\end{eqnarray}
Obviously, the information update process has a very simple form and
only the parameter $\nu$ is affected by the new observation
$n_{k+1}$. The posterior density
(\ref{Chapter_combining_formula:numberposteriori}) does not change
its type every time new data arrive and hence, is easily calculated.

 The moments of a GIG are not available in a closed form through
elementary functions but can be expressed in terms of Bessel
functions. In particular, the posterior expected number of losses is
\begin{equation}
\label{Chapter_combining_expectedvalue} \mathrm{E}[\Lambda|
\vec{N}=\vec{n}, \vec{\Delta}=\vec{\delta}] =
\sqrt{\frac{\phi}{\omega}}\frac{K_{\nu + 2}(2 \sqrt{\omega
\phi})}{K_{\nu+1}(2 \sqrt{\omega \phi})}.
\end{equation}

\noindent The mode of a GIG has a simple expression  that gives the
posterior mode
\begin{equation} \textnormal{mode}(\Lambda| \vec{N}=\vec{n}, \vec{\Delta}=\vec{\delta}) = \frac{1}{2 \omega}
(\nu + \sqrt{\nu^2 + 4 \omega \phi}).
\end{equation}
\noindent It can be used as an alternative point estimator instead
of the mean. Also, the mode of a GIG differs only slightly from the
expected value for large $|\nu|$. A full asymptotic interpretation
of the Bayesian estimator (\ref{Chapter_combining_expectedvalue})
can be found Lambrigger et al \cite{LaShWu07} that shows the model
behaves as we would expect and require in practice.

\subsection{Numerical example}\label{Chapter_combining_example:frequeny} A simple example, taken
from  Lambrigger et al \cite[example 3.7]{LaShWu07}, illustrates the
above methodology combining three data sources. It also extends
numerical example from Section
\ref{Chapter_combining_PoissonGamma_example}, where two data sources
are combined using classical Bayesian inference approach. Assume
that:

\begin{itemize}
\item External data
(for example, provided by external databases or regulator) estimate
the intensity of the loss frequency (i.e. the Poisson parameter
$\Lambda$), which has a prior gamma distribution $\Lambda \sim Gamma
(\alpha_0,\beta_0)$, as $\mathrm{E}[\Lambda]=\alpha_0 \beta_0 = 0.5$
and $\Pr[0.25 \leq \Lambda \leq 0.75]= 2/3$. Then, the parameters of
the prior  are $\alpha_0 \approx 3.407$ and $\beta_0 \approx 0.147$;
see Section \ref{Chapter_combining_PoissonGamma_example}.

\item One expert gives an estimate of the intensity as $\delta=0.7$. For simplicity,
we consider in this example one single expert only and hence, the
coefficient of variation is not estimated using
(\ref{Chapter_combining_eq:xiestimate}), but given a priori (e.g. by
the regulator): $\mathrm{Vco}[\Delta|\Lambda] =
\sqrt{\mathrm{Var}[\Delta|\Lambda]} / \mathrm{E}[\Delta|\Lambda] =
0.5$, i.e. $\xi=4$.

\item The observations of the annual number of losses $n_1,n_2,\ldots$ are sampled from $Poisson(0.6)$ and are the same
as in Section \ref{Chapter_combining_PoissonGamma_example}.

\end{itemize}

\noindent This means that a priori we have a frequency parameter
distributed as $Gamma(\alpha_0,\beta_0)$ with mean $\alpha_0 \beta_0
= 0.5$. The true value of the parameter $\lambda$ for this risk in a
bank is $0.6$, that is, it does worse than the average institution.
However, our expert has an even worse opinion of his institution,
namely $\delta = 0.7$. Now, we compare:
\begin{figure}[!htbp]
\begin{center}
    \includegraphics[scale=0.8]{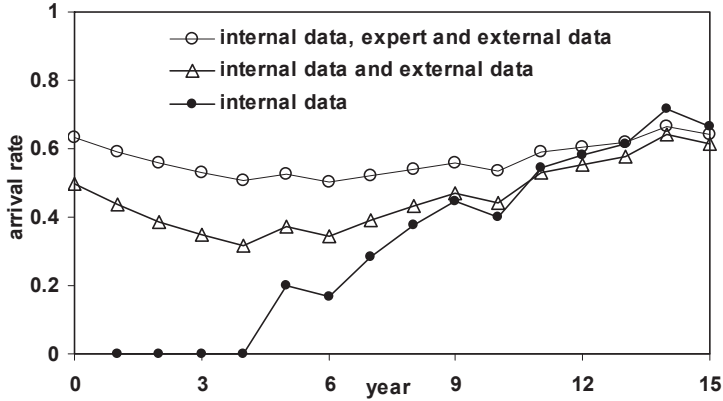}
    \caption{\small{$(\circ)$ The Bayes estimate $\widehat{\lambda}_k^{(3)}$, $k=1,\ldots,15$,
    combines the internal data simulated from $Poisson(0.6)$, external data giving
    $\mathrm{E}[\Lambda]=0.5$, and expert opinion $\delta=0.7$.
    It is compared with the Bayes estimate $\widehat{\lambda}_k^{(2)}$ $(\triangle)$, that combines external data and internal
    data,
    and the classical maximum likelihood estimate $\widehat{\lambda}^{\mathrm{MLE}}_k$
     $(\bullet)$. See Example \ref{Chapter_combining_example:frequeny} for
     details.}}
    \label{Chapter_combining_fig:bayesclaimnumber1}
\end{center}
\end{figure}

\begin{itemize}

\item the pure maximum likelihood estimate
$\widehat{\lambda}_k^{\mbox{\tiny{MLE}}} = \frac{1}{k}
\sum_{i=1}^{k} n_i;$

\item the Bayesian estimate (\ref{Chapter_combining_PoissonFreq_Expected_eq}),
$ \widehat{\lambda}_k^{(2)} =
\mathrm{E}[\Lambda|N_1=n_1,\ldots,N_k=n_k]$, without expert opinion;

\item the Bayesian estimate derived in formula
(\ref{Chapter_combining_expectedvalue}) $ \widehat{\lambda}_k^{(3)}
= \mathrm{E}[\Lambda|N_1=n_1,\ldots,N_k=n_k,\Delta=\delta]$, that
combines internal data and expert opinions with the prior.
\end{itemize}

\noindent The results are plotted in Figure
\ref{Chapter_combining_fig:bayesclaimnumber1}. The estimator
$\widehat{\lambda}_k^{(3)}$ shows a much more stable behaviour
around the true value $\lambda=0.6$, due to the use of the prior
information (market data) and the expert opinions. Given adequate
expert opinions, $\widehat{\lambda}_k^{(3)}$ clearly outperforms the
other estimators, particularly if only a few data points are
available.

One could think that this is only the case when the experts'
estimates are appropriate. However, even if experts fairly under-
(or over-) estimate the true parameter $\lambda$, the method
presented here performs better for our dataset than the other
mentioned methods, when a few data points are available. The above
example yields a typical picture observed in numerical experiments
that demonstrates that the Bayes estimator
(\ref{Chapter_combining_expectedvalue}) is often more suitable and
stable than maximum likelihood estimators based on internal data
only. Note that in this example the prior distribution as well as
the expert opinion do not change over time. However, as soon as new
information is available or when new risk management tools are in
place, the corresponding parameters may be easily adjusted.

\begin{remark} In this section, we considered the situation where
${\Lambda }$ is the same for all years $t=1,2,\ldots$ . However, in
general, the evolution of ${\Lambda }_{t}$, can be modelled as
having deterministic (trend, seasonality) and stochastic components,
the case when ${\Lambda }_{t}$ is purely stochastic and distributed
according to a gamma distribution is considered in Peters, et al
\cite{PeShWu09b}.
\end{remark}

\subsection{Lognormal Model for Severities}
\label{Chapter_combining_LNsev3dataSources} In general, one can use
the methodology summarised by equation
(\ref{Chapter_combining_aposterioridistribution}) to develop a model
combining external data, internal data and expert opinion for
estimation of the severity. For illustration purposes, this section
considers the lognormal severity model.

Consider modelling severities $X_{1}, \ldots, X_{K},\ldots\;$ using
the lognormal distribution $\mathcal{LN}(\mu,\sigma)$, where
$\vec{X}=(X_{1}, \ldots, X_{K})^\prime$ are the losses over past $T$
years. Here, we take an approach considered in Section
\ref{Chapter_coimbining_LognormalUnknownMu_sec}, where $\mu$ is
unknown and $\sigma$ is known. The unknown $\mu$ is treated under
the Bayesian approach as a random variable $\Theta_\mu$. Then
combining external data, internal data and expert opinions can be
accomplished using the following model.

\begin{model_assumptions}[Lognormal-normal-normal]
\label{Chapter_combining_model:subexp} Let us assume the following
severity model for a risk cell in one bank:
\begin{itemize}
\item[a)] Let $\Theta_\mu \sim \mathcal{N} (\mu_0, \sigma_0)$ be a
normally distributed random variable with parameters $\mu_0,
\sigma_0$, which are estimated from (external) market data, i.e.
$\pi(\vec{\theta})$ in
(\ref{Chapter_combining_aposterioridistribution}) is the density of
$\mathcal{N}(\mu_0,\sigma_0)$.
\item[b)] Given $\Theta_\mu=\mu$, the losses $X_1,X_2,\ldots\;$ are
conditionally independent with a common lognormal distribution:
$X_{k}{\sim} \mathcal{LN}(\mu,\sigma)$, where $\sigma$ is assumed
known. That is, $f_1(\cdot|\mu)$ in
(\ref{Chapter_combining_aposterioridistribution}) corresponds to the
density of a $\mathcal{LN}(\mu,\sigma)$ distribution.
\item[c)] The financial company has $M$ experts with opinions
$\Delta_m$, $1 \leq m \leq M$, about $\Theta_\mu$. Given
$\Theta_\mu=\mu$, $\Delta_m$ and $X_k$ are independent for all $m$
and $k$, and $\Delta_1,\ldots,\Delta_M$ are independent with a
common normal distribution: $\Delta_m {\sim} \mathcal{N} (\mu,\xi)$,
where $\xi$ is a parameter estimated using expert opinion data. That
is, $f_2(\cdot|\mu)$ corresponds to the density of a
$\mathcal{N}(\mu,\xi)$ distribution.
\end{itemize}
\end{model_assumptions}
\begin{remark}

~

\begin{itemize}
\item For $M \geq 2$, the parameter $\xi$ can be estimated by the standard deviation of $\delta_m$:
\begin{equation} \widehat{\xi} = \left(\frac{1}{M - 1} \sum_{m=1}^{M}
(\delta_m - \overline{\delta})^2 \right)^{1/2}.
\end{equation}
\item The hyper-parameters $\mu_0$ and $\sigma_0$ are estimated from market data, for example, by maximum likelihood estimation or by the method of moments.
\item In practice one often uses an ad-hoc estimate for $\sigma$, which usually is
based on expert opinion only. However, one could think of a Bayesian
approach for $\sigma$, but then an analytical formula for the
posterior distribution in general does not exist and the posterior
 needs then to be calculated numerically, for example, by MCMC methods.
\end{itemize}
\end{remark}
Under Model Assumptions \ref{Chapter_combining_model:subexp}, the
posterior density is given by
\begin{eqnarray}
{\pi}(\mu | \vec{x}, \vec{\delta}) &\propto& \frac{1}{\sigma_0
\sqrt{2 \pi}} \exp \left( - \frac{(\mu-\mu_0)^2}{2 \sigma_0^2}
\right)
\prod_{k=1}^{K}  \frac{1}{\sigma \sqrt{2 \pi}} \exp \left( - \frac{(\ln x_{k}-\mu)^2}{2 \sigma^2} \right)\nonumber\\
&& \prod_{m=1}^{M}  \frac{1}{\xi \sqrt{2 \pi}} \exp \left( - \frac{(\delta_m-\mu)^2}{2 \xi^2} \right)\nonumber\\
&\propto & \exp \left[-\left(\frac{(\mu-\mu_0)^2}{2 \sigma_0^2} +
\sum_{k=1}^{K} \frac{(\ln x_{k}-\mu)^2}{2\sigma^2}
+ \sum_{m=1}^{M} \frac{(\delta_m-\mu)^2}{2\xi^2}  \right) \right] \nonumber \\
&\propto& \exp \left[ -\frac{(\mu - \widehat{\mu})^2}{2
\widehat{\sigma}^2} \right],
\end{eqnarray}
with \begin{equation*} \widehat{\sigma}^2 = \left(
\frac{1}{\sigma_0^2} + \frac{K}{\sigma^2} + \frac{M}{\xi^2}
\right)^{-1},
\end{equation*} and
\begin{equation*} \widehat{\mu} =
\widehat{\sigma}^2 \times \left( \frac{\mu_0}{\sigma_0^2} +
\frac{1}{\sigma^2} \sum_{k=1}^{K} \ln x_{k} + \frac{1}{\xi^2}
\sum_{m=1}^{M} \delta_m \right).
\end{equation*}
In summary, we derived the following theorem (also see Lambrigger et
al \cite{LaShWu07}). That is, the posterior distribution of
$\Theta_\mu$, given loss information $\vec{X}=\vec{x}$ and expert
opinion $\vec{\Delta}=\vec{\delta}$, is a normal distribution
$\mathcal{N}(\widehat{\mu},\widehat{\sigma})$ with
\begin{equation*}
\widehat{\sigma}^2 = \left( \frac{1}{\sigma_0^2} +
\frac{K}{\sigma^2} + \frac{M}{\xi^2} \right)^{-1}
\end{equation*}
and
\begin{equation}
\label{Chapter_combining_eq:subexpmean} \widehat{\mu} =
\mathrm{E}[\Theta_\mu| \vec{X}=\vec{x}, \vec{\Delta}=\vec{\delta}] =
\omega_1 \mu_0 + \omega_2 \overline{\ln x} + \omega_3
\overline{\delta},
\end{equation}
where $\overline{\ln x} = \frac{1}{K} \sum_{k=1}^{K} \ln x_{k}$ and
the credibility weights are
$$\omega_1 = \widehat{\sigma}^2 /
\sigma_0^2,\; \omega_2 = \widehat{\sigma}^2 K / \sigma^2,\; \omega_3
= \widehat{\sigma}^2 M / \xi^2.$$

This yields a natural interpretation. The more credible the
information, the higher is the credibility weight in
(\ref{Chapter_combining_eq:subexpmean}) -- as expected from an
appropriate model for combining internal observations, relevant
external data and expert opinions.

\section{Nonparametric Bayesian
approach}\label{nonparametricBayesian_sec} Typically, under the
Bayesian approach, we assume that there is unknown distribution
underlying observations $x_1,\ldots,x_n$ and this distribution is
parametrized by $\bm\theta$. Then we place a prior distribution on
the parameter $\bm\theta$ and try to infer the posterior of
$\bm\theta$ given observations $x_1,\ldots,x_n$. Under the
nonparametric approach, we do not make assumption that underlying
loss process generating distribution is parametric; we put prior on
the distribution directly and find the posterior of the distribution
given data which is combining of the prior with empirical data
distribution.

One of the most popular Bayesian nonparametric models is based on
Dirichlet process introduced in Ferguson \cite{Ferguson73}. The
Dirichlet process represents a probability distribution of the
probability distributions. It can be specified in terms of a base
distribution $H(x)$ and a scalar concentration parameter $\alpha>0$
and denoted as $DP(\alpha, H)$. For example, assume that we model
severity distribution $F(x)$ which is unknown and modelled as random
at each point $x$ using $DP(\alpha, H)$. Then, the mean value of
$F(x)$ is the base distribution $H(x)$ and variance of $F(x)$ is
$H(x)(1-H(x))/(\alpha+1)$. That is, as the concentration parameter
$\alpha$ increases, the true distribution is getting closer to the
base distribution $H(x)$. Each draw from Dirichlet process is a
distribution function and for $x_1<x_2<\cdots<x_k$, the distribution
of
$$F(x_1),F(x_2)-F(x_1),\ldots,1-F(x_k)$$
is a $k+1$ multivariate Dirichlet distribution
$$Dir(\alpha
H(x_1),\alpha(H(x_2)-H(x_1)),\ldots,\alpha(1-H(x_k)))$$ formally
defined as follows.

\begin{definition}[Dirichlet distribution]\index{Dirichlet distribution}\index{Dirichlet process}
A d-variate Dirichlet distribution is denoted as
$Dir(\alpha_1,\alpha_2,\ldots,\alpha_d)$, where $\alpha_i>0$. The
random vector $(Q_1,Q_2,\ldots,Q_d)$ has a Dirichlet distribution if
its density function is
\begin{equation}
f(q_1,q_2,\ldots,q_{d-1})=\frac{\Gamma(\alpha_1+\cdots+\alpha_d)}{\prod_{i=1}^d\Gamma(\alpha_i)}\prod_{i=1}^d
q_i^{\alpha_i-1},
\end{equation}
where $q_i>0$ and $q_1+\cdots+q_d=1$.
\end{definition}

There are several formal definitions of Dirichlet processes; for
detailed description  see Ghosh and Ramamoorthi
\cite{GhoshRamamoorthi2003}. For the purposes of this book, here we
just present few important results that can be easily adopted for
OpRisk. In particular, the $i$th marginal distribution of
$Dir(\alpha_1,\ldots,\alpha_d)$ is
$Beta(\alpha_i,\alpha_0-\alpha_i)$, where
$\alpha_0=\alpha_1+\cdots+\alpha_d$. Thus the marginal distribution
of the Dirichlet process $DP(\alpha, H)$ is beta distribution
$F(x)\sim Beta(\alpha H(x),\alpha(1-H(x)))$, i.e. explicitly it has
the Beta density
\begin{equation}\label{DirichletMarginalBetaBounds_eq}
\Pr[F(x)\in dy]=\frac{\Gamma(\alpha)}{\Gamma(\alpha
H(x))\Gamma(\alpha(1-H(x)))}y^{\alpha
H(x)}(1-y)^{\alpha(1-H(x))-1}dy,
\end{equation}
where $\Gamma(\cdot)$ is a gamma function.

If the prior distribution for $F(x)$ is $DP(\alpha, H)$, then after
observing $x_1,\ldots, x_n$, the posterior for $F(x)$ is
\begin{equation}
DP\left(\alpha+n,
\frac{\alpha}{\alpha+n}H(x)+\frac{n}{\alpha+n}\frac{1}{n}\sum
1_{x_i\le x}\right).
\end{equation}

In other words, Dirichlet process is a conjugate prior with respect
to empirical sample distribution; in posterior, our unknown
distribution $F(x)$ will have updated concentration parameter
$\alpha+n$ and updated base distribution
\begin{equation}\label{Dirichlet_basedistributionUpdate_eq}
\widetilde{H}(x)=\frac{\alpha}{\alpha+n}H(x)+\frac{n}{\alpha+n}\frac{1}{n}\sum
1_{x_i\le x},
\end{equation}
which is a weighted sum of the prior base distribution and empirical
distribution with the weights ${\alpha}/(\alpha+n)$ and
$n/(\alpha+n)$ respectively. The modeller can choose $H(x)$ as an
expert opinion on distribution $F(x)$, then posterior estimate of
the $F(x)$ after observing data $x_1,\ldots,x_n$ will be given by
$\widetilde{H}(x)$ in (\ref{Dirichlet_basedistributionUpdate_eq}).

\begin{remark}
~
\begin{itemize}
\item As new data are collected, the
posterior distribution converges to the empirical distribution that
itself converges to the true distribution of $F(x)$.

\item The larger value of
$\alpha$, the less impact new data will have on the posterior
estimate of $F(x)$; if $\alpha=0$, the posterior distribution will
simply be the empirical distribution of the data.

\item The concentration parameter $\alpha$ can be interpreted as an``effective
sample size� associated with the prior estimate. In assigning the
value of c, the modeler should attempt to quantify the level of
information contained in the scenario estimates, as measured by the
equivalent amount of data that would provide a similar level of
confidence. The modeller can also estimate $\alpha$ from a likely
interval range of severities or frequencies (i.e. from the variance
of the possible distribution). Cope \cite{Cope2012} suggests that
given the rarity of the scenarios considered, the assigned value of
$\alpha$ will likely be low, often less than ten and possibly as low
as one.
\end{itemize}
\end{remark}

~

\begin{figure}[!htbp]
\centerline{\includegraphics[scale=0.7]{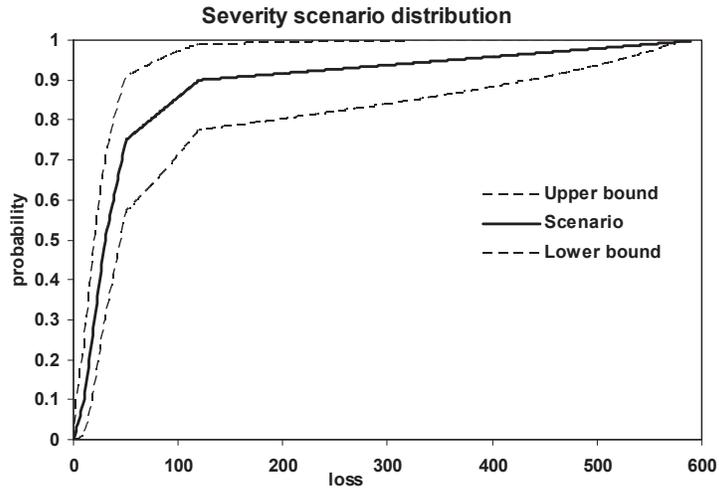}}
\caption{\small{Dirichlet marginal bounds for scenario severity
distribution.}} \label{DirichletBounds_fig}
\end{figure}

\begin{figure}[!htbp]
\centerline{\includegraphics[scale=0.7]{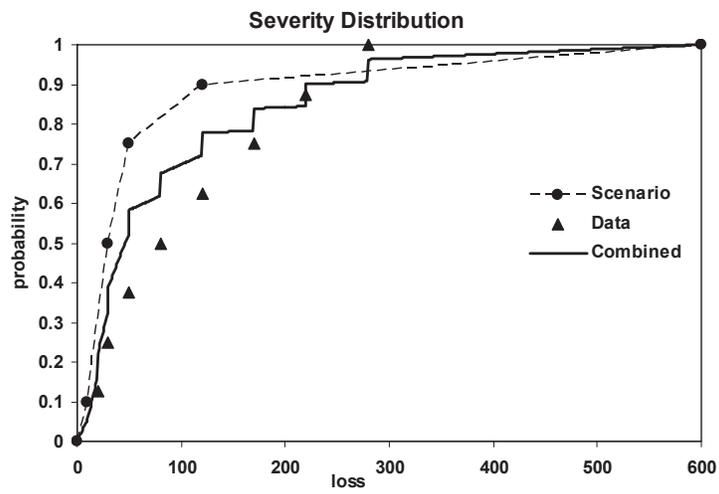}}
\caption{\small{Combining scenario severity distribution with
empirical distribution of the observed data.}}
\label{DirichletCombiningExample_fig}
\end{figure}

\noindent\textbf{Numerical Example.} Assume that expert provides
estimates in USD millions for a risk severity as follows. If loss
occurs, then the probability to exceed 10, 30, 50 and 120 are 0.9,
0.5, 0.25 and 0.1 respectively, and the maximum possible loss is USD
600 million. That is, probability distribution $H(x)$ at points (0,
10, 30, 50, 120, 600) is (0, 0.1, 0.5, 0.75, 0.9, 1). It is
presented in Figure \ref{DirichletBounds_fig} with linear
interpolation between specified distribution points. If we choose
the prior for the unknown severity distribution $F(x)$ as
$DP(\alpha, H(x))$ with concentration parameter $\alpha=10$, then
expected value for $F(x)$ from the prior is $H(x)$ and bounds for
$F(x)$ for each $x$ can be calculated from the marginal beta
distribution (\ref{DirichletMarginalBetaBounds_eq}). For example,
the lower and upper bounds in Figure \ref{DirichletBounds_fig}
correspond to 0.1 and 0.9 quantiles of the beta distribution
$Beta(\alpha H(x),\alpha(1-H(x)))$, i.e. will contain the true value
of $F(x)$ with probability 0.8 for each $x$. Now, assume that we
observe the actual losses 20, 30, 50, 80, 120, 170, 220, and 280 all
in USD million. The posterior mean of $F(x)$ combining scenario and
data is easily calculated using
(\ref{Dirichlet_basedistributionUpdate_eq}) and presented in Figure
\ref{DirichletCombiningExample_fig} along with the empirical data
and scenario distribution.


\section{Combining using Dempster-Shafer
structures}\label{DempsetShafer_sec} Often risk assessment includes
situations where there is little information on which to evaluate a
probability or information is nonspecific, ambiguous, or
conflicting. In this case one can work with bounds on probability.
For example, this idea has been developed in Walley and Fine
\cite{WalleyFine1982}, Berleant \cite{Berleant1993} and there are
suggestions that the idea has its roots from Boole \cite{Boole1854}.
Williamson and Downs \cite{WilliamsonDowns1990} introduced
interval-type bounds on cumulative distribution functions called
"probability boxes" or "p-boxes". They also described algorithms to
compute arithmetic operations (addition, subtraction, multiplication
and division) on pairs of p-boxes.

The method of reasoning with uncertain information known as
Dempster-Shafer theory of evidence was suggested in Dempster
\cite{Dempster1967,Dempster68} and Shafer \cite{Shafer76}. A special
rule to combine the evidence from different sources was formulated
in Dempster \cite{Dempster68}; it is somewhat controversial and
there are many modifications to the rule such as in Yager
\cite{Yager1986,Yager1987}.

For a good summary on the methods for obtaining Dempster-Shafer
structures and ``p-boxes", and aggregation methods handling a
conflict between the objects from different sources, see Ferson
\emph{et al} \cite{FeKrGiMySe03}. The use of p-boxes and
Dempster-Shafer structures in risk analyses offers many significant
advantages over a traditional probabilistic approach. Ferson
\emph{et al} \cite{FeKrGiMySe03} lists the following practical
problems faced by analysts that can be resolved using these methods:
 imprecisely specified distributions,
 poorly known or even unknown dependencies,
 non-negligible measurement uncertainty,
 non-detects or other censoring in measurements,
 small sample size,
 inconsistency in the quality of input data,
 model uncertainty, and
 non-stationarity (non-constant distributions).

It is emphasized in Walley \cite{Walley1991} that the use of
imprecise probabilities does not require one to assume the actual
existence of any underlying distribution function. This approach
could be useful in risk analyses even when the underlying stochastic
processes are nonstationary or could never, even in principle, be
identified to precise distribution functions. Oberkampf et al
\cite{OberkampfEtAl2001} and Oberkampf \cite{Oberkampf2005}
demonstrated how the theory could be used to model uncertainty in
engineering applications of risk analysis stressing that the use of
p-boxes and Dempster-Shafer structures in risk analyses offers many
significant advantages over a traditional probabilistic approach.

These features are certainly attractive for OpRisk, especially for
combining expert opinions, and were applied for OpRisk in Sakalo and
Delasey \cite{SakaloDelasey2011}. At the same time, some writers
consider these methods as unnecessary elaboration that can be
handled within the Bayesian paradigm through Baysian robustness
(section 4.7 in Berger \cite{Berger85}). Also, it might be difficult
to justify application of Dempster's rule (or its other versions) to
combine statistical bounds for empirical data distribution with
exact bounds for expert opinions.

\subsection{Dempster-Shafer structures and p-boxes}
 A Dempster-Shafer
structure on the real line is similar to a discrete distribution
except that the locations where the probability mass resides are
sets of real values (\emph{focal elements}) rather than points. The
correspondence of probability masses associated with the focal
elements is called the basic probability assignment. This is
analogous to the probability mass function for an ordinary discrete
probability distribution. Unlike a discrete probability distribution
on the real line, where the mass is concentrated at distinct points,
the focal elements of a Dempster-Shafer structure may overlap one
another, and this is the fundamental difference that distinguishes
Dempster-Shafer theory from traditional probability theory.
Dempster-Shafer theory has been widely studied in computer science
and artificial intelligence, but has never achieved complete
acceptance among probabilists and traditional statisticians, even
though it can be rigorously interpreted as classical probability
theory in a topologically coarser space.

\begin{definition}[Dempster-Shafer
structure]\label{Dempster-Shafer structure_def} A finite
Dempster-Shafer structure on the real line $\mathbb{R}$ is
probability assignment, which is a mapping
$$m : 2^{\mathbb{R}}\rightarrow [0; 1],$$
where $m(\emptyset) = 0$; $m(a_i) = p_i$ for focal elements
$a_i\subseteq {\mathbb{R}}$, $i = 1, 2, \ldots, n$; and $m(D) = 0$
whenever $D\neq a_i$ for all $i$, such that $0 < p_i$ and
$p_1+\cdots+p_n = 1$.
\end{definition}
For convenience, we will assume that the focal elements $a_i$ are
closed intervals $[x_i,y_i]$. Then implementation of a
Dempster-Shafer structure will require $3n$ numbers; one for each
$p_i$; and $x_i$ and $y_i$ for each corresponding focal element.

\begin{remark}
Note that $ 2^{\mathbb{R}}$ denotes a power set. The power set of a
set $\mathbb{S}$ is the set of all subsets of $\mathbb{S}$ including
the empty set $\emptyset$ and $\mathbb{S}$ itself. If $\mathbb{S}$
is a finite set with $K$ elements then the number of elements in its
power set is $2^K$. For example, if $\mathbb{S}$ is the set
$\{x,y\}$, then the power set is $\{\emptyset,x,y,\{x,y\}\}$.
\end{remark}

The upper and lower probability bounds can be defined for
Dempster-Shafer structure. These are called \emph{plausibility} and
\emph{belief} functions defined as follows.

\begin{definition}[Plausibility function]
The plausibility function corresponding to a Dempster-Shafer
structure $m(A)$ is the sum of all masses associated with sets that
overlap with or merely touch the set $b\subseteq{\mathbb{R}}$
$$Pls(b)=\sum_{a_i\cap b\neq\emptyset}m(a_i),$$
which is the sum over $i$ such that $a_i\cap b\neq\emptyset$.
\end{definition}
\begin{definition}[Belief function]
The belief function corresponding to a Dempster-Shafer structure
$m(A)$ is the sum of all masses associated with sets that are
subsets of
 $b\subseteq{\mathbb{R}}$
$$Bel(b)=\sum_{a_i\subseteq b}m(a_i),$$
which is the sum over $i$ such that $a_i\subseteq b$.
\end{definition}
Obviously, $Bel(b)\le Pls(b)$. Also, if one of the structures
(either Dempster-Shafer structure, or $Bel$ or $Pls$) is known then
the other two can be calculated. Considering sets of all real
numbers less than or equal to $z$, it is easy to get upper and lower
bounds for a probability distribution of a random real-valued
quantity characterized by a finite Dempster-Shafer structure.

Consider Dempster-Shafer structure with focal elements that are
closed intervals $[x_i,y_i]$. We can specify it by listing focal
elements the focal elements and their associated probability masses
$p_i$ as $\{([x_1, y_1], p_1), ([x_2, y_2], p_2),\ldots, ([x_n,
y_n], p_n)\}$. Then the left bound (cumulative plausibility
function) and the right bound (cumulative belief function) are
\begin{equation}\label{CumPlausibilityFunc_eq}
{F}^{U}(z)=\sum_{x_i\le z}p_i;\quad {F}^{L}(z)=\sum_{y_i\le z}p_i.
\end{equation}
respectively. These functions are non-decreasing and right
continuous functions from real numbers onto the interval $[0,1]$ and
$F^L(z)\le F^U(z)$, i.e. proper distribution functions. They define
the so-called p-box $[F^L(z),F^U(z)]$ that can be defined without
any reference to Dempster-Shafer structure.
\begin{definition}[probability box or p-box]
p-box is a set of all probability distributions $F(x)$ such that
$F^L\le F(x)\le F^U(x)$, where $F^L(x)$ and $F^U(x)$ are
nondecreasing functions from the real line into $[0,1]$. It is
denoted as $[F^L,F^U]$.
\end{definition}
That is, we say that $[F^L,F^U]$ is a p-box of a random variable $X$
whose distribution $F(x)$ is unknown except that $F^L\le F(x)\le
F^U(x)$.

\begin{example}\label{DempsterShaferStructure_example1}
Consider the following Dempster-Shafer structure with three focal
elements that have the same probability $1/3$, i.e.
$$
\mbox{Structure A} = \left\{\begin{array}{cc}
                     {[x_1=5,y_1=20]}; & p_1=1/3 \\
                     {[x_2=10,y_2=25]}; & p_2=1/3 \\
                     {[x_3=15,y_3=30]}; & p_3=1/3
                   \end{array}
  \right.
$$
\noindent Plausibility and belief functions are easily calculated
using (\ref{CumPlausibilityFunc_eq}) respectively and presented by
structure A in Figure \ref{DempsterShaferStructureExample_fig}.
\end{example}

\subsection{Dempster's rule}\label{DempsterRule_sec}
The central method in the Dempster-Shafer theory is Dempster's rule
for combining evidence (Shafer \cite{Shafer76}; Dempster
\cite{Dempster1967}). In some situations, this rule produces
counterintuitive results and various alternative versions of the
rule have been suggested such as Yager \cite{Yager1987}. In this
section, we briefly describe only the original Dempster's rule which
is used to combine evidence obtained from two or more independent
sources for the same quantity in question (e.g. expert opinions
about a specific risk). A considerably more extensive review of this
literature is available in Sentz and Ferson \cite{SentzSentz2002}.

\begin{definition}[Dempster's rule]
The combination of two independent Dempster-Shafer structures
$m_1(A)$ and $m_2(B)$ with focal elements $a_i$ and $b_j$
respectively is another Dempster-Shafer structure with probability
assignment
\begin{equation}\label{DempsterRule_eq}
m(\emptyset)=0;\quad m(c)=\frac{1}{1-{\mathds{K}}}\sum_{a_i\cap
b_j=c} m_1(a_i)m_2(b_j)\quad\mbox{for}\;\; c\neq \emptyset,
\end{equation}
i.e. the sum over all $i$ and $j$ such that intersection of $a_i$
and $b_j$ is equal to $c$, where
\begin{equation}\label{DempsterRule_conflict_eq}
{\mathds{K}}=\sum_{a_i\cap b_j=\emptyset}m_1(a_i)m_2(b_j)
\end{equation}
is the mass associated with the conflict present in the combining
evidence.
\end{definition}

\begin{example}\label{DempsterShaferCombining_example1}
Consider two independent Dempster-Shafer structures A and B with
focal elements $a_i$ and $b_j$ respectively
$$
\mbox{Structure A} = \left\{\begin{array}{cc}
                     {[5,20]},\frac{1}{3} &  \\
                     {[10,25]},\frac{1}{3} &  \\
                     {[15,30]},\frac{1}{3} &
                   \end{array}
  \right.\quad \mbox{and} \quad
\mbox{Structure B} = \left\{\begin{array}{cc}
                     {[10,25]},\frac{1}{3} & \\
                     {[15,30]},\frac{1}{3} &  \\
                     {[22,35]},\frac{1}{3} &
                   \end{array}
  \right.
$$
The only combination of focal elements between these two structures
that has no intersection is $a_1=[5,20]$  with $b_3=[22,35]$. Thus
the conflict of information in (\ref{DempsterRule_conflict_eq}) is
${\mathds{K}}=\frac{1}{3}\frac{1}{3}=\frac{1}{9}$. Using Dempster
rule (\ref{DempsterRule_eq}) to combine structures A and B, we
obtain the following structure C:
$$ \left\{              ({[10,20]},\frac{1}{8});
                     ({[15,20]},\frac{1}{8});
                     ({[10,25]},\frac{1}{8};
                     ({[15,25]},\frac{1}{4});
                     ({[22,25]},\frac{1}{8});
                     ({[15,30]},\frac{1}{8});
                     ({[22,30]},\frac{1}{8})\right\}.
$$
Note that intersection $c_4=[15,25]$ is produced by two
combinations: $a_2$ with $b_2$; and $a_3$ with $b_1$. Thus $c_4$ has
probability
$(\frac{1}{3}\frac{1}{3}+\frac{1}{3}\frac{1}{3})/(1-{\mathds{K}})=1/4$
while all other elements of structure $C$ are produced by one
combination and have probability
$\frac{1}{3}\frac{1}{3}/(1-{\mathds{K}})=\frac{1}{8}$ each.
Plausibility and belief functions of all structures are easily
calculated using (\ref{CumPlausibilityFunc_eq}  and presented in
Figure \ref{DempsterShaferStructureExample_fig} for all structures.
elements.
\begin{figure}[!htbp]
\centerline{\includegraphics[scale=0.65]{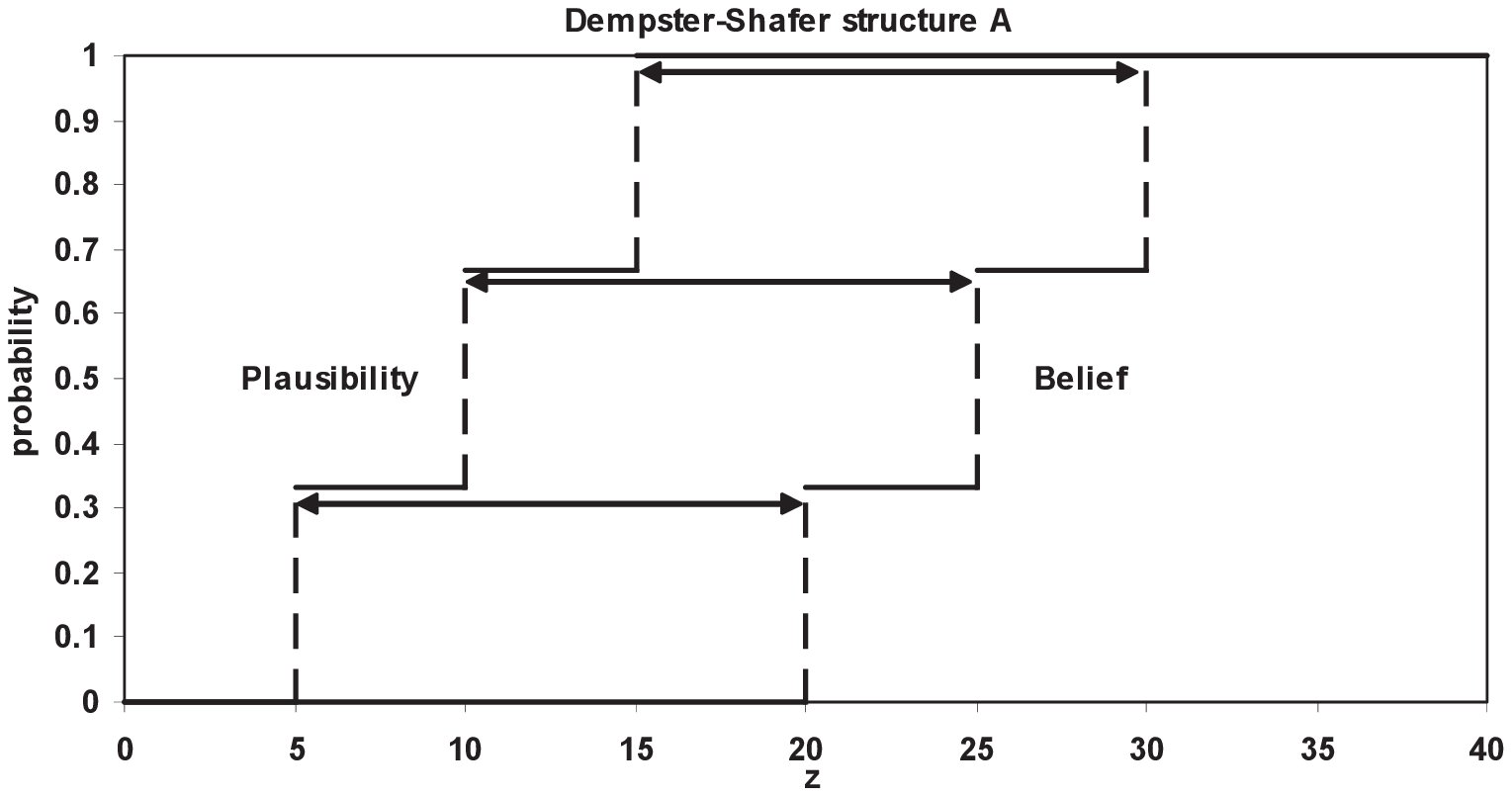}}
\centerline{\includegraphics[scale=0.65]{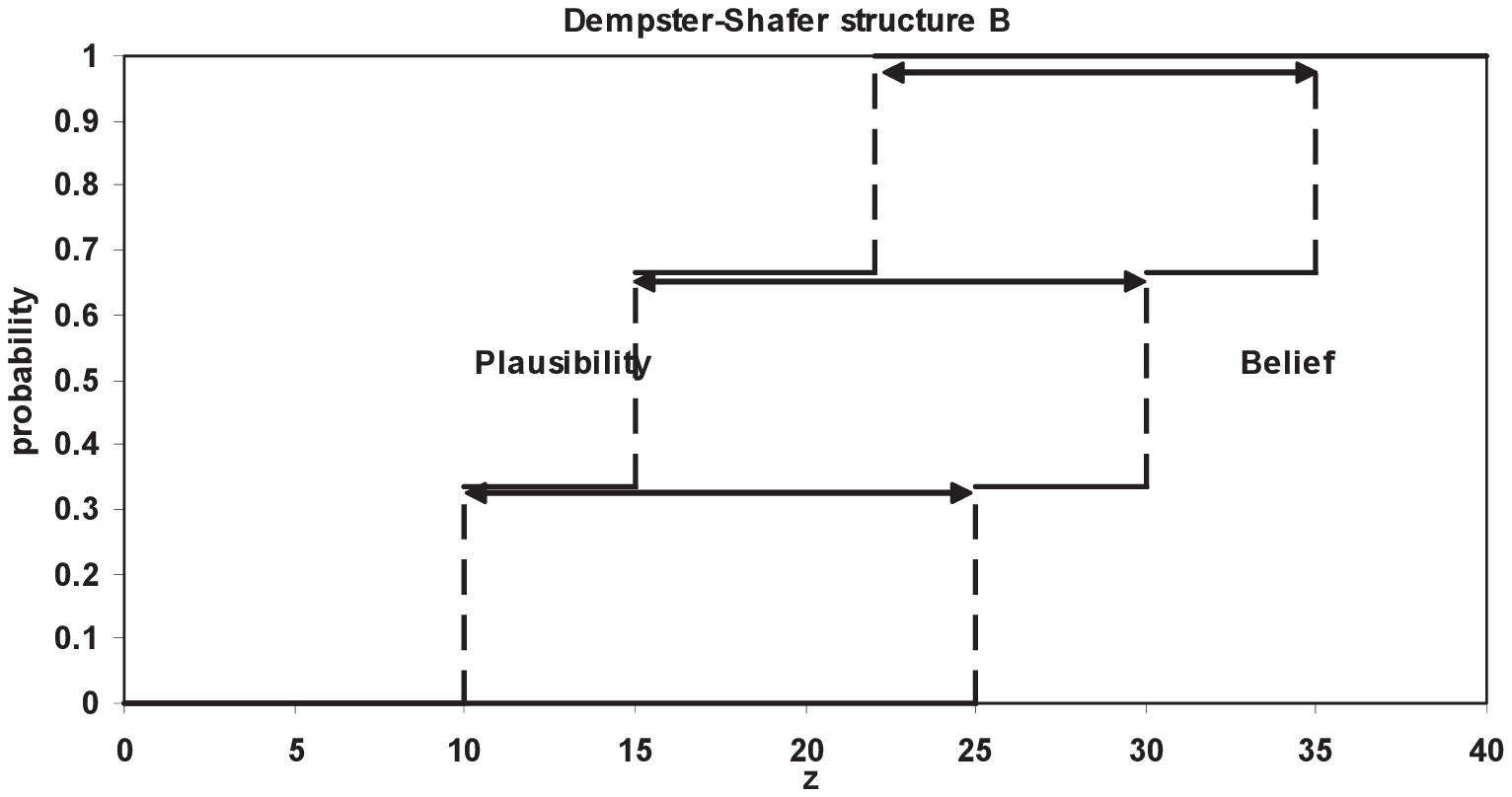}}
\centerline{\includegraphics[scale=0.65]{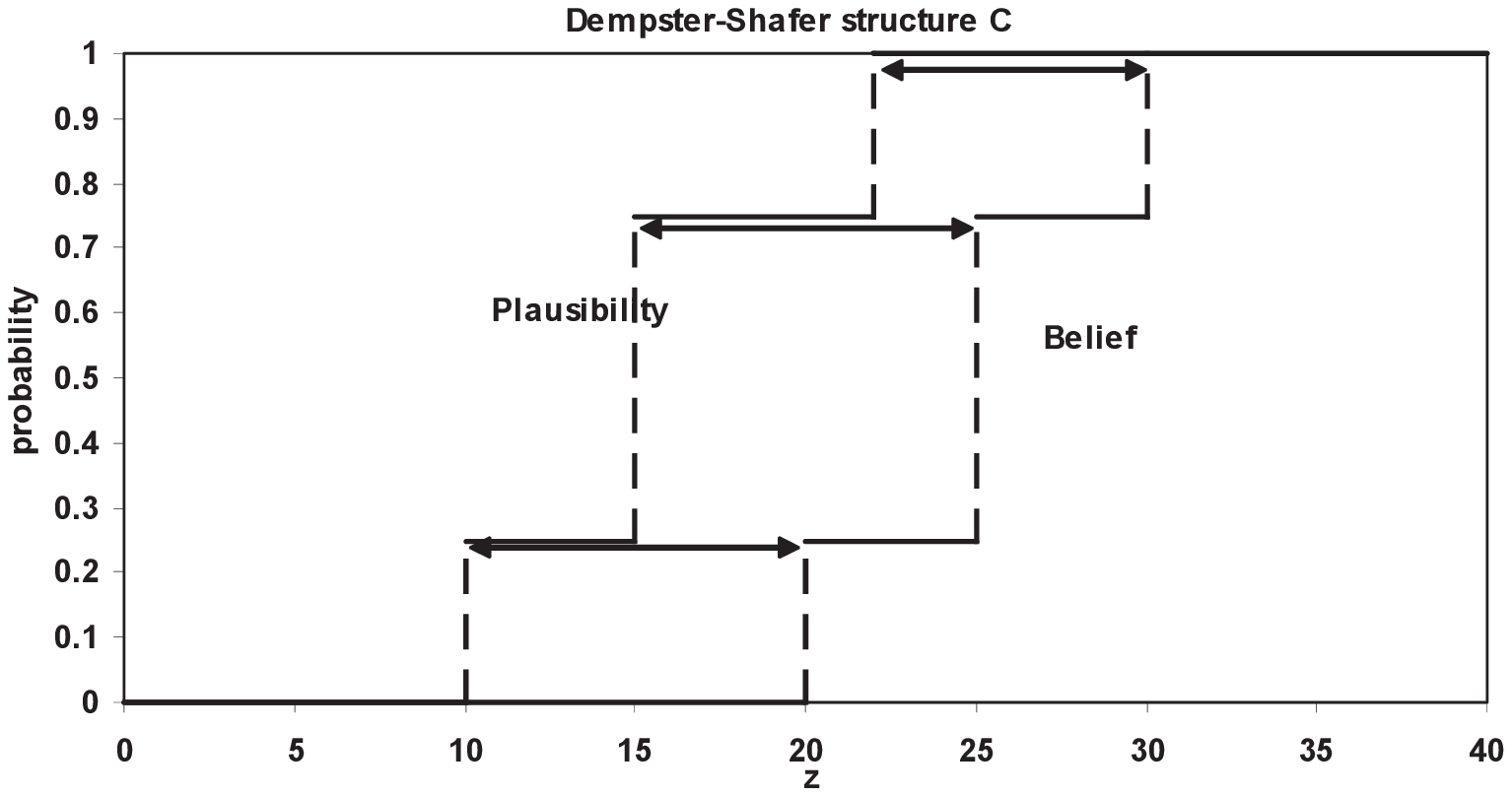}}
\caption{\small{Plausibility and belief functions for
Dempster-Shafer structures in example in Section
\ref{DempsterRule_sec}. Focal elements of the structure are
indicated by arrows. Structure C is a result of combining strcutures
A and B via Dempster's rule.}}
\label{DempsterShaferStructureExample_fig}
\end{figure}
\end{example}

\subsection{Intersection method}
If the estimates to be aggregated represent claims that the quantity
has to be within some limits, then \emph{intersection method} is
perhaps the most natural kind of aggregation. The idea is simply to
use the smallest region that all estimates agree. For example, if we
know for sure that a true value of the quantity $a$ is within the
interval $x = [1, 3]$, and we also know from another source of
evidence, that $a$ is also within the interval $y = [2, 4]$, then we
may conclude that $a$ is certainly within the interval $x\cap y =
[2, 3]$.

The most general definition of intersection can be specified in
terms of probability boxes. If there are $K$ p-boxes $F_1 = [F_1^L
,F_1^U ],\ldots,F_K = [F_K^L ,F_K^U ]$, then their intersection is a
p-box $[F^L,F^U]$, where
\begin{equation}
F^U=\min(F_1^U,\ldots,F_K^U),\quad F^L=\max(F_1^L,\ldots,F_K^L)
\end{equation}

\noindent if $F^L(x)\le F^U(x)$ for all $x$. This operation is used
when the analyst is highly confident that each of multiple p-boxes
encloses the distribution of the quantity in question. This
formulation extends to Dempster-Shafer structures easily. The
cumulative plausibility and belief functions of such structures form
p-boxes.

Despite its several desirable properties, the intersection has only
limited application for aggregation in OpRisk because it requires a
very strong assumption that the individual estimates are each
absolutely correct. It is certainly not recommended to the cases if
any of the experts might be wrong. In practice, wrong opinions can
be more typical than correct ones. For more detailed discussion and
examples, see Ferson \emph{et al} \cite{FeKrGiMySe03}.

\subsection{Envelope method}
In the previous section on aggregation via intersection, it is
assumed that all the estimates to be aggregated are completely
reliable. If the analyst is confident only that at least one of the
 estimates encloses the quantity, but does not know which estimate,
the method of \emph{enveloping} can be used to aggregate the
estimates into one reliable characterization. In general, when the
estimates to be aggregated represent claims about the true value of
a quantity and these estimates have uncertain reliability,
enveloping is often an appropriate aggregation method. The idea is
to identify the region where any estimate might be possible as the
aggregation result. In particular, if one expert says that the value
is $1$ and another expert says that it is $2$, we might decide to
use the interval $[1,2]$ as the aggregated estimate. If there are
$K$ p-boxes $F_1 = [F_1^L ,F_1^U ],\ldots,F_K = [F_K^L ,F_K^U ]$,
then their envelope is defined to be a p-box $[F^L,F^U]$ where
\begin{equation}
F^U=\max(F_1^U,\ldots,F_K^U),\quad F^L=\min(F_1^L,\ldots,F_K^L)
\end{equation}
This operation is always defined. It is used when the analyst knows
that at least one of multiple p-boxes describes the distribution of
the quantity in question. This formulation extends to
Dempster-Shafer structures easily. The cumulative plausibility and
belief functions of such structures form p-boxes. The result of
aggregating these p-boxes can then be translated back into a
Dempster-Shafer structure by canonical discretization. However,
enveloping  is sensitive to claims of general ignorance. This means
that if only one  expert provides an inconclusive opinion, it will
determine the result of the aggregation. The overall result of
enveloping will be as broad as the broadest input. The naive
approach to omit any inconclusive estimates before calculating the
envelope will not be sufficient in practice because any estimate
that is not meaningless but just very wide can swamp all other
estimates. Again, for more detailed discussion, the reader is
referred to Ferson \emph{et al} \cite{FeKrGiMySe03}.

\subsection{Bounds for empirical data distribution}
P-boxes and Dempster-Shafer structures can be constructed for
empirical data using distribution free bounds around an empirical
distribution function  (Kolmogorov
\cite{Kolmogorov1941,Kolmogorov1933}; Smirnov \cite{Smirnov1939}).
Similar to the confidence intervals around a single number, these
are bounds on a statistical distribution as a whole. As the number
of samples increases, these confidence limits would converge to the
empirical distribution function. Given independent samples
$X_1,\ldots X_n$ from unknown continuous distribution $F(x)$, the
empirical distribution of the data is
$$
F_n(x)=\frac{1}{n}\sum_1^n 1_{X_i\le x}.
$$
The lower and upper bounds (referred to as Kolmogorov-Smirnov
bounds) for the distribution $F(x)$ can be calculated as
\begin{eqnarray}\label{KSbounds_eq}
F_n^L=\max(0,F_n(x) - D(\alpha,n));\quad F_n^U=\min(1,F_n(x) +
D(\alpha,n)),
\end{eqnarray}
where $D(\alpha,n)$ is a critical value for the one-sample
Kolmogorov-Smirnov statistic $D_n$ at the confidence level
$100(1-\alpha)\%$ and sample size $n$, i.e.
$$
\Pr[D_n\le D(\alpha,n)]=1-\alpha, \quad\mbox{where} \quad
D_n=\sup_{x}\left|F_n(x)-F(x)\right|.
$$
The tabulated values for $D(\alpha, n)$ as well as a numerical
approximations can be found in Miller \cite{Miller1956}. For
example, for sample size $n=10$ and $\alpha=0.05$ (i.e. $95\%$
confidence level), $D(\alpha,n)=0.40925$. Note that typically,
Kolmogorov-Smirnov statistics $D_n$ is used for goodness-of-fit
testing to compare a sample with a reference probability
distribution. The null hypothesis that sample is from $F(x)$ is
rejected at level $\alpha$ if $D_n$ exceeds critical value
$D(\alpha,n)$.

\begin{figure}[!htbp]
\centerline{\includegraphics[scale=0.7]{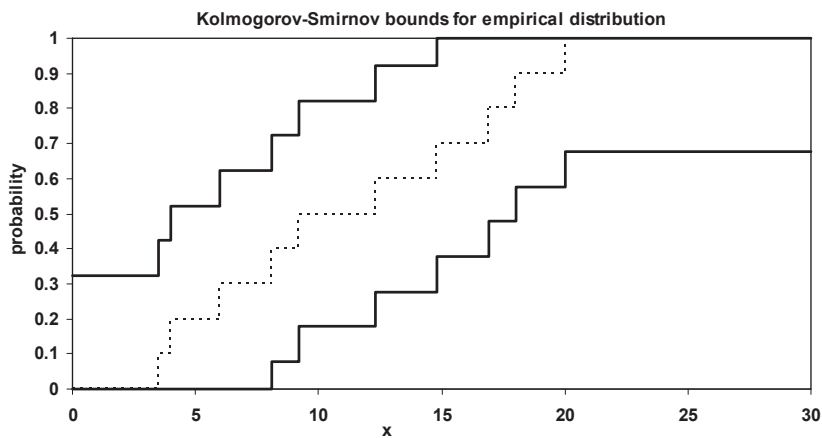}}
\caption{\small{Kolmogorov-Smirnov bounds for empirical
distribution; for details see Example \ref{KSbounds_example}.}}
\label{KSboundsExample_fig}
\end{figure}

Theoretically, the left tail of the KS upper limit extends to
negative infinity. But, of course, the smallest possible value might
be limited by other considerations. For instance, there might be a
theoretical lower limit at zero. If so, we could use this fact to
truncate the upper (left) bound at zero. The right tail of the lower
limit likewise extends to positive infinity. Sometimes it may be
reasonable to select some value at which to truncate the largest
value of a quantity too.
\begin{example}\label{KSbounds_example}
Assume that we have the following iid samples
$$(3.5; 4; 6; 8.1; 9.2; 12.3;
14.8; 16.9; 18; 20)$$ Also assume that the lower bound for samples
is zero and the upper bound is 30. Then Kolmogorov-Smirnov bounds at
$80\%$ confidence are calculated using (\ref{KSbounds_eq}) and
presented in Figure \ref{KSboundsExample_fig}.
\end{example}
The Kolmogorov-Smirnov bounds make no distributional assumptions,
but they do require that the samples are independent and identically
distributed. In practice, an independence assumption is sometimes
hard to justify. Kolmogorov-Smirnov bounds are widely used in
probability theory and risk analyses, for instance as a way to
express the reliability of the results of a simulation.

Formally, the Kolmogorov-Smirnov test is valid for continuous
distribution functions. Also, in the discrete case,
Kolmogorov-Smirnov bounds are conservative, i.e. these bounds can be
used in the case of discrete distributions but may not represent
best possible bounds.

The confidence value $\alpha$ should be chosen such that the analyst
believes the p-box contains the true distribution. The same
hypothesis must also be assumed for the construction of the p-box
from expert estimates. However, note that a p-box defined by
Kolmogorov-Smirnov confidence limits is fundamentally different from
the sure bounds. The Kolmogorov-Smirnov bounds are not certain
bounds but statistical ones. The associated statistical statement is
that $95\%$ (or whatever is specified by $\alpha$) of the time the
true distribution will be within the bounds. It is not completely
clear how to combine the Kolmogorov-Smirnov p-box with the expert
specified p-box; the choices of the upper limit and confidence level
$\alpha$ for Kolmogorov-Smirnov bounds can be problematic.

\section{Conclusions}
In this paper we reviewed several methods suggested in the
literature for combining different data sources required for the LDA
under Basel II requirements. We emphasized that Bayesian methods can
be well suited for modeling OpRisk. In particular, Bayesian
framework is convenient to combine different data sources (internal
data, external data and expert opinions) and to account for the
relevant uncertainties. There are many other methodological
challenges in the LDA implementation such as modelling dependence,
data truncation and estimation which are under the hot debate in the
literature; for a recent review, the reader is referred to
Shevchenko \cite{shevchenko2011modelling}.

\footnotesize{
\bibliography{HandbookOpRisk_bibliographyGWPandPS_Feb}
\bibliographystyle{wileyj}
}
\end{document}